\documentclass[aps,prx,showpacs,twocolumn,reprint]{revtex4-2}

\usepackage{qcircuit}
\usepackage[dvips]{graphicx}
\usepackage{amsmath,amssymb,amsthm,mathrsfs,amsfonts,dsfont}
\usepackage{subfigure, epsfig}
\usepackage{braket}
\usepackage{hyperref}
\usepackage{bm}
\usepackage{enumerate}
\usepackage{color}
\usepackage{graphicx}
\usepackage{pgf}

\usepackage{amsthm}

\newtheorem{theorem}{Theorem}

\newcommand{\tr}{\mathrm{Tr}}

\newcommand{\qf}{\mathbf{F}_Q}

\newcommand{\var}{\mathrm{Var}}

\begin{document}

% --------------------  TITLE  --------------------

\title{Quantum Analytic Descent}

% ------------  AUTHORS AND AFFILIATIONS ----------
\author{B\'alint Koczor}
\email{balint.koczor@materials.ox.ac.uk}
\affiliation{Department of Materials, University of Oxford, Parks Road, Oxford OX1 3PH, United Kingdom}
\author{Simon C. Benjamin}
\affiliation{Department of Materials, University of Oxford, Parks Road, Oxford OX1 3PH, United Kingdom}

% --------------------  ABSTRACT  --------------------

\begin{abstract}
Variational algorithms have particular relevance for near-term quantum computers but require non-trivial parameter optimisations. Here we propose Analytic Descent: Given that the energy landscape must have a certain simple form in the local region around any reference point, it can be efficiently approximated in its entirety by a classical model -- we support these observations with rigorous, complexity-theoretic arguments. One can classically analyse this approximate function in order to directly `jump' to the (estimated) minimum, before determining a more refined function if necessary. We derive an optimal measurement strategy and generally prove that the asymptotic resource cost of a ‘jump’ corresponds to only a single gradient vector evaluation.
\end{abstract}

\maketitle

\section{Introduction}

Quantum devices have already been announced
whose behaviour cannot be simulated using classical computers with practical levels
of resource~\cite{GoogleSupremacy,PhysRevLett.127.180502,PhysRevLett.127.180501,ebadi2021quantum}. In this era, quantum computers may have the potential
to perform useful tasks of value.
The early machines will not have a comprehensive solution to
accumulating noise~\cite{preskill2018quantum}, and therefore it is a considerable and fascinating
challenge to achieve a valuable function despite the imperfections.
One very promising class of approaches
are generically called quantum variational algorithms in which one seeks to make use of a
quantum circuit of (presumably) relatively low depth~\cite{farhi2014quantum,peruzzo2014variational,endo2020hybrid, cerezo2020variationalreview, bharti2021noisy},
by adjusting the function it performs to tune it to the desired task. 

Typically a simple-to-prepare reference state (such as all-zero) is passed into a quantum circuit,
called the ansatz circuit, within which there are numerous parametrised gates. The idea exists in many
variants, both theoretical and experimental~\cite{farhi2014quantum,peruzzo2014variational,wang2015quantum,PRXH2,PhysRevA.95.020501,mcclean2016theory,
	PhysRevLett.118.100503,Li2017,PhysRevX.8.011021,Santagatieaap9646,kandala2017hardware,kandala2018extending,
	PhysRevX.8.031022,romero2018strategies,higgott2018variational,mcclean2017hybrid,kokail2018self,sharma2020noise,cerezo2020variational},
refer also to the review articles \cite{endo2020hybrid, cerezo2020variationalreview, bharti2021noisy}.
In a typical implementation, each gate implements a unitary which is therefore also parametrised;
for example
\begin{equation}\label{pauli-gate}
\exp(-i \theta \sigma_x /2),
\end{equation}
where $\sigma_x$ is the Pauli $X$ operator acting on a given qubit, and $\theta$ is the classical parameter.
For a suitably-chosen ansatz circuit and an appropriate number of independently parametrised gates, the emerging
state (also called the ansatz state) may be very complex -- while inevitably being restricted to a small
proportion of the exponentially large Hilbert space.
A given problem, for example the challenge of finding the ground state of some molecule of interest,
is encoded by deriving a Hamiltonian $\mathcal{H}$ whose ground state represents an acceptable solution.
This is of course a non-trivial challenge in itself for many systems of interest. Importantly, this challenge
is not decoupled from the task of selecting a suitable ansatz circuit, or that of choosing the initial parameters
for that circuit. Assuming that all these tasks have been appropriately performed then the hope is that there
exists some set of parameters, to be discovered, for which the ansatz state emerging from the circuit is indeed
(acceptably close to) the desired solution state. The problem then becomes one of parameter search -- there might
easily be hundreds of parameters, so that techniques from classical optimisation are very relevant to the
prospects of successfully finding the proper configuration.

A popular choice is gradient descent; in the basic form of this method one evaluates the gradient of
energy $\langle \mathcal{H} \rangle$  with respect to each of the ansatz parameters. One then takes a
`small step' in the direction of steepest gradient descent, and re-evaluates the gradient. Numerous adaptions
are of course possible, ranging from varying the size of the step through to more advanced protocols for obtaining
a valid direction of progress~\cite{van2021measurement}.

Although gradient descent and its more advanced variants,
such as natural gradient \cite{quantumnatgrad, koczor2019quantum,samimagtime} are a popular choice,
they have their limitations and costs. Determining the energy in quantum chemistry or in recompilation problems must be performed
to a very high accuracy to be useful (e.g. chemical accuracy, equivalent to 3 or 4 decimal places~\cite{ourReview})
while finding the minimum of the energy landscape very precisely is generally expensive
due to shot noise \cite{arrasmith2020operator,van2021measurement,romero2018strategies}.

In the present work we study an alternative method of particular relevance in
the latter stages of a QVA when we begin to approach the minimum of the cost function:
Using an ansatz circuit within which gates correspond to Pauli strings (a universal construction), we observe that the cost function
i.e. the expected energy of the output state with respect to the problem Hamiltonian,
will necessarily have certain simple properties in the local region around any reference point.
Exploiting this knowledge, we sample from the ansatz circuit to determine an analytic function to the near-minimum region.
Given this function, we can descend towards the minimum classically and then take a `large jump' (as
compared to the small incremental steps taken in generic gradient descent) direct to that point.
If necessary we then repeat the procedure of refining the analytic function and jumping again, until
we reach a point satisfactorily close to the true minimum.
We derive an optimal measurement strategy whereby we occasionally collect further samples during a descent
	to reduce shot noise in our classical approximations.
In numerical simulations of this approach
we find that a single jump can be equivalent of thousands of steps of generic gradient descent.

\section{Expanding the ansatz circuit}
Quantum gates generated by Pauli strings have only two distinct eigenvalues,
and consequently as we vary the parameter $\theta$ associated
with such a gate,
the corresponding slice of the energy surface is simply of the form
$a + b\cos\theta +c \sin\theta$ for some $a,b,c \in \mathbb{R}$. For further discussion refer to
\cite{PhysRevResearch.2.043158,parrish2019jacobi,ostaszewski2019quantum,paramshift,vidal2018calculus,schuld2020effect}.

It immediately follows from the above property that the Fourier spectrum of
the full energy surface is determined by $3^\nu$ coefficients, where $\nu$
is the number of parameters. Despite the very simple
structure of such functions, determining them classically
is intractable. Nevertheless, previous works proposed that
the \emph{exact} energy surface can be reconstructed for a classically tractable
number of parameters, e.g., $\nu  = 1, 2$, while freezing other parameters
and thereby sequentially optimising the surface using a classical computer
\cite{PhysRevResearch.2.043158,parrish2019jacobi,ostaszewski2019quantum}.
Here we make the fundamental observation that one could efficiently
obtain -- by estimating \emph{at most} a quadratic number of terms --
a good \emph{classical approximation} of the full energy surface (and its full gradient vector)
that is valid in the vicinity of any reference point. We support these
observations with rigorous, complexity-theoretic arguments and with an optimal measurement 
strategy. Let us introduce our model.

We define an ansatz circuit as a CPTP mapping, and in particular,
 as a product of individual gate operations that we write in terms
 of their superoperators as
\begin{equation}\label{fullansatz}
\Phi(\underline{\theta}):= \Phi_\nu(\theta_\nu) \dots \Phi_2(\theta_2) \Phi_1(\theta_1).
\end{equation}	
Here $\Phi_k(\theta_k)$ are parameterised quantum gates, such as in Eq.~\eqref{pauli-gate}.
We focus on quantum gates which are generated by Pauli strings
as (approximately) unitary operators $\Phi_k(\theta_k) \rho \approx U \rho U^\dagger $
with $U=\exp(- i \theta_k P_k/2)$. Here $P_k$ are products of single-qubit
Pauli operators as $P_k \in \{\mathrm{Id}, \sigma_x, \sigma_y, \sigma_z \}^{\otimes N}$.
For any such ansatz circuit, we can expand every gate into the following form. First, let us fix
$\theta_0$ and express the continuous dependence of the quantum gates on
the angle $\theta$ around the fixed $\theta_0$ as
\begin{equation}\label{single-gate}
\Phi_k(\theta_0 + \theta) = 
a(\theta) \Phi_{ak}
+ b(\theta) \Phi_{bk}
+ c(\theta) \Phi_{ck},
\end{equation}
where $a(\theta), b(\theta) = 1\pm\cos(\theta)$ and $c(\theta) = \sin(\theta)/2$
are simple Fourier components.
The transformations can be specified as $\Phi_{ak} = \Phi_k(\theta_0)$,
and via parameter shifts as $\Phi_{bk} = \Phi_k(\theta_0+\pi/2) - \Phi_k(\theta_0-\pi/2)$
and $\Phi_{ck}  =\Phi_k(\theta_0+\pi)$. Note that these transformations are discrete in nature,
and implicitly depend on the constant $\theta_{0}$ which we have fixed as a reference point.
Refer to Appendix~\ref{singlePauliSec} for more details.

We now expand the full ansatz circuit from Eq.~\eqref{fullansatz} into the above form
assuming that all gates are generated by Pauli strings via Eq.~\eqref{single-gate}.
We again fix $\underline{\theta}_0$ and express the continuous
dependence on $\underline{\theta}$ around this reference point in parameter space as
\begin{equation}\label{ansatz-approx}
\Phi(\underline{\theta}_0 {+} \underline{\theta})
= 
\prod_{k=1}^\nu
[a(\theta_k) \Phi_{ak}
+ b(\theta_k) \Phi_{bk}
+ c(\theta_k) \Phi_{ck}].
\end{equation}	
The above product can be expanded into a sum of $3^\nu$ terms, which 
is classically intractable. Nevertheless, in the following
we aim to approximate this mapping via a polynomial number
of \emph{important} summands and discard the remaining, less important terms.
In particular, we introduce $\delta := \lVert \underline{\theta}\rVert_\infty$,
which denotes the absolute largest entry in the parameter vector.
We will now expand the above
quantum circuit into a quadratic number of terms in $\nu$ which introduces an
error $\mathcal{O}(\sin^3 \delta)$.

We derive the explicit form of this approximate mapping in
Appendix~\ref{fullansatz-section} as
\begin{align} \label{full-circuit}
\Phi(\underline{\theta}) = A(\underline{\theta}) \Phi^{(A)} &+ \sum_{k=1}^\nu
[B_k(\underline{\theta}) \Phi^{(B)}_k + C_k(\underline{\theta}) \Phi^{(C)}_k] \\
&+\sum_{l>k}^\nu [ D_{kl}(\underline{\theta}) \Phi^{(D)}_{kl}] + \mathcal{O}(\sin^3 \delta). \nonumber
\end{align}
Here $A, B_k, C_k, D_{kl} : \mathbb{R}^\nu \mapsto \mathbb{R}$ are
multivariate functions -- in fact, monomials in $a(\theta), b(\theta)$ and $c(\theta)$ --
and they multiply the discrete mappings as, e.g., $A(\underline{\theta}) \Phi^{(A)}$.
As such, these monomials are products of simple univariate trigonometric functions and they
completely absorb the continuous dependence on the parameters $\underline{\theta}$.

Our derivation of Eq.~\eqref{full-circuit} is detailed in Appendix~\ref{fullansatz-section} and
relies on the following 3 steps. First, we substitute the explicit forms of the single-variate
trigonometric functions $a(\theta_k), b(\theta_k)$ and $c(\theta_k)$ into Eq.~\eqref{ansatz-approx}.
Second, we expand the resulting product into a sum of $3^\nu$ terms. Third, we discard
all contributions that contain a product of 3 or more $\sin\theta_k$ terms thereby introducing
an error $\mathcal{O}(\sin^3 \delta)$.
We could similarly approximate the mapping via a sum of $\mathcal{O}(\nu^3)$ terms up to an
error $\mathcal{O}(\sin^4\delta)$ or beyond.

Our multivariate trigonometric series has the significant advantage that it can
capture some global features in contrast to local Taylor expansions, 
	for example,
	the approximation is exact along single parameter slices and respects symmetries
	of the objective function, such as its periodicity.
	In particular, while the error term is generally upper bounded by the `pessimistic'
	monomial $\mathrm{const} \times \delta^3$
	just like in the case of a Taylor expansion, the actual error can be significantly below this bound
	and, e.g., can be zero along single parameter slices. Moreover the constant prefactor in the above
	upper bound can be significantly smaller than in case of a Taylor expansion, refer to
	Appendix~\ref{hessianSection} and to Appendix~\ref{symmetry}.

%-------------------------
\begin{figure*}[tb]
	\begin{centering}
		\includegraphics[width=0.8\textwidth]{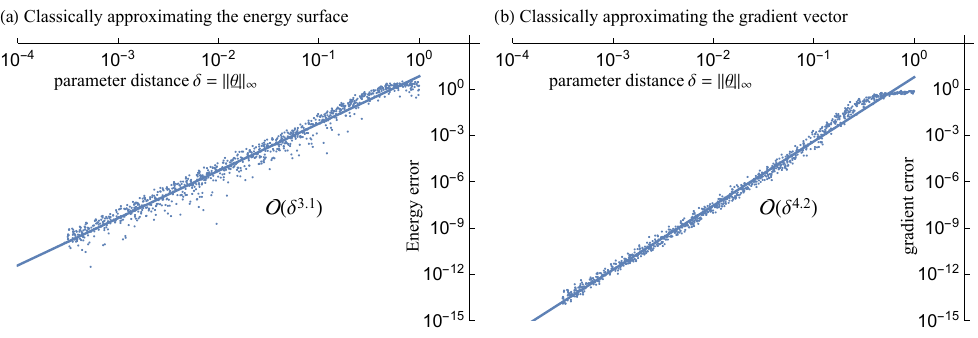}
		\caption{
			Error of our trigonometric-series approximation of the entire energy surface (a) and gradient vector (b) as a function
			of the distance $\delta$ from the reference point $\underline{\theta}_0$ of our model, where
			$\delta=\lVert \underline{\theta} \rVert_\infty$ is the absolute largest parameter $\theta_k$. As long as $\delta$ is small, we can classically approximate the
			gradient vector and use it in an analytic descent optimisation. 
			The approximation error of the gradient vector is computed as the similarity measure $1-f$,
			refer to text. We used a 12-qubit spin-ring Hamiltonian as in Fig.~\ref{fig2}(b) and an 84-parameter ansatz
			circuit, and included the empirical scaling of the errors as $\mathcal{O}(\delta^{3.1})$
			and $\mathcal{O}(\delta^{4.2})$.
			\label{fig1}
		}
	\end{centering}
\end{figure*}
%-------------------------

\section{Classically computing the entire energy surface}
A large class of potential 
applications in the context of variational quantum algorithms assume a target function
that corresponds to \emph{linear mappings} of the form
$E(\underline{\theta}) := \tr[\mathcal{H} \,  \Phi(\underline{\theta})\rho_0]$, that
can be used to model, e.g., the expected energy of a physical system when $\mathcal{H}$ is
a Hamiltonian \cite{endo2020hybrid, cerezo2020variationalreview, bharti2021noisy}.

Using our expansion in Eq.~\eqref{full-circuit}, we can express the entire energy surface
explicitly as
\begin{align} \label{full-energy}
E(\underline{\theta}) = A(\underline{\theta}) E^{(A)} &+ \sum_{k=1}^\nu
[B_k(\underline{\theta}) E^{(B)}_k + C_k(\underline{\theta}) E^{(C)}_k] \\
&+\sum_{l>k}^\nu [ D_{kl}(\underline{\theta}) E^{(D)}_{kl}] + \mathcal{O}(\sin^3 \delta). \nonumber
\end{align}
Here $E^{(A)},E^{(B)}_k,E^{(C)}_k, E^{(D)}_{kl} \in \mathbb{R}$ can be reconstructed by estimating  
the energy expectation value at discrete points in parameter space using a quantum device. For example,
$E^{(A)} = \tr[ \mathcal{H} \, \Phi^{(A)} \rho_0 ] = E(\underline{\theta}_0)$ is just the
energy at the fixed point $\underline{\theta}_0$ and $E^{(C)}_k$ is obtained similarly
by shifting the $k^{\text{th}}$ parameter by $\pi$.
As such, our classical approximation algorithm depends on a quadratic number
	of coefficients that can be fully determined by querying energy expectation values.
Indeed, error mitigation techniques are applicable
\cite{endo2020hybrid, koczor2021exponential,koczor2021dominant, huggins2020virtual}.

Let us here briefly summarise our derivation of Eq.~\eqref{full-energy} from Appendix~\ref{energySection}.
	We first apply both sides of Eq.~\eqref{full-circuit} to our reference state $\rho_0$, i.e.,
	on the left-hand side we obtain the exact, continuously parametrised quantum state
	$\Phi(\underline{\theta})\rho_0$ while 
	on the right-hand side we obtain an approximation to it in
	terms of the discrete mappings, such as $\Phi^{(B)}_k \rho_0$. We finally obtain Eq.~\eqref{full-energy}
	by computing quantum-mechanical expected
	values via the linear mapping $\tr[\mathcal{H} \,  \cdot]$, for example, we obtain
	the coefficients as $E^{(B)}_k =  \tr[\mathcal{H} \, \Phi^{(B)}_k \rho_0 ]$
	which expresses the well-know parameter shift rule as
	$E(\underline{\theta} + \tfrac{1}{2}\pi \underline{v}_k) -   E(\underline{\theta}-\tfrac{1}{2}\pi \underline{v}_k)$.

Fig.~\ref{fig1}(a) shows approximation errors
obtained from a simulated ansatz circuit of 12 qubits
as a function of the
absolute largest entry in the parameter vector given by the norm $\delta =\lVert \underline{\theta} \rVert_\infty$.
We computed the approximate energy via Eq.~\eqref{full-energy} at 1000 randomly generated points 
in parameter space in the vicinity of our reference point $\underline{\theta}_0$, close to the global optimum.
Fig.~\ref{fig1}(a) confirms the error scaling $\mathcal{O}( \delta^3 )$ and illustrates
that the error is smaller than $10^{-3}$ as long as the parameter vector norm
$\lVert \underline{\theta} \rVert_\infty$ is smaller than $0.1$. 
We further remark that in
Appendix~\ref{symmetry}
we derive exact and approximate symmetries of the
energy function around local minima; the objective function is approximately
reflection symmetric via $E(\underline{\theta}) \approx E(-\underline{\theta})$
and exactly reflection symmetric along slices $\theta_k$.

\section{Classically computing the gradient}
We derive the full analytic gradient of the approximate energy surface from Eq.~\eqref{full-energy} in
Appendix~\ref{gradientSection}
and propose an efficient classical algorithm for computing it for a given input $\underline{\theta}$
in Appendix~\ref{classical-alg-sec}.
This has a classical computational complexity of $\mathcal{O}(\nu^3)$.
We simulate an ansatz circuit of 12 qubits in Fig.~\ref{fig1}(b) and compute the
approximation error of the analytically calculated gradient vector.
We quantify this error using the similarity measure as the
normalised scalar product $f = \langle \underline{\tilde{g}} | \underline{g}\rangle /
(\lVert \underline{\tilde{g}} \rVert  \lVert \underline{g} \rVert)$, between the
exact $\underline{g} $ and the approximate $\underline{\tilde{g}}$ gradient vectors. We plot $1-f$
in Fig.~\ref{fig1}(b), and conclude that our approximation is very good and
that our error measure scales with the parameter vector norm
in fourth order as $1-f = \mathcal{O}( \delta^4 )$.

We aim to use this gradient vector in a classical optimisation routine,
but we first have to take into account shot noise:
When using a quantum
device to estimate the coefficients in Eq.~\eqref{full-energy}, one needs to collect a
large number of samples in order to sufficiently reduce the statistical uncertainty in those estimates.
This uncertainty is quantified by the variances as, e.g., $\var[ E^{(B)}_k]$ when
estimating the coefficient $E^{(B)}_k$. 
As such, we want to determine the gradient vector to a fixed precision
as the expected Euclidean distance $\epsilon^2:= \langle \lVert\Delta g\rVert^2 \rangle = \sum_{k=1}^\nu \var[ \partial_m E(\underline{\theta}) ]$
for which we derive the following error propagation formula
\begin{align}	\label{grad_variance_main}
	\epsilon^2	=&  \mathcal{A}(\underline{\theta}) \var[ E^{(A)}]  
	+ \sum_{k=1}^\nu \mathcal{B}_k(\underline{\theta}) \, \var[ E^{(B)}_k]\\
	&+ \sum_{k=1}^\nu \mathcal{C}_k(\underline{\theta}) \, \var[ E^{(C)}_k]
	+	\sum_{l>k} \mathcal{D}_{kl} (\underline{\theta}) \,  \var[E^{(D)}_{kl}].
		\nonumber
\end{align}
Here $\mathcal{A}, \mathcal{B}_k, \mathcal{C}_k,\mathcal{D}_{kl} : \mathbb{R}^\nu \mapsto \mathbb{R}$
are trigonometric polynomials that depend on the parameters $\underline{\theta}$
and we can efficiently compute these \cite{gitcode}.
Note that the above statistical uncertainties are directly proportional to single-shot variances
of estimating energy expectation values as, e.g., $\var[E^{(A)}]  = \var[E(\underline{\theta}_0)]$,
and advanced estimation techniques can be applied \cite{Crawford2019}.

We analytically derive an optimal
measurement strategy in Theorem~\ref{theo:opt_dist} which does not require us
to determine the coefficients,
such as $E^{(B)}_k$, in order to predict their measurement costs but only their
variances which is relatively cheap.
As such, using a small overhead in quantum resources, our classical algorithm takes
an input parameter vector $\underline{\theta}$ and it exactly determines how many
measurements need to be assigned to estimating the individual coefficients.
Most importantly, when we are close to our reference point almost all
measurements are assigned to the coefficients $E^{(B)}_k$ which guarantees that the cost of
our approach is comparable to a single iteration of gradient descent.

For this reason, we prove the following approximate upper bound of the full measurement costs
	in Theorem~\ref{relative_cost} relative
to determining a single gradient vector as
\begin{equation}\label{meas_upb}
	N/N_{grad}   \leq   [ 1 +  S (\sqrt{2} + \nu) \delta \, ]^2 + \mathcal{O}(\delta^2 ) + \mathcal{O}(\nu \delta^3 ).
\end{equation}
Here $S$ is the ratio of minimal and maximal single-shot variances due to estimating the energy
$E(\underline{\theta})$ at different points $\underline{\theta}$
while $\nu$ is the number of ansatz parameters and $\delta$ is the distance from our
reference point $\underline{\theta}_0$.	
In our proof in Appendix~\ref{theo2sec} we expand the \emph{exact}
variance propagation formula from Eq.~\eqref{grad_variance_main} and 
obtain Eq.~\eqref{meas_upb} by keeping only the leading terms in $\delta$
and upper bounding the single-shot variances.

Our upper bound in Eq.~\eqref{meas_upb} ensures us of the following:
a) Initialising analytic descent in the reference point $\underline{\theta}_{0}$ costs
exactly the same as determining a single gradient vector;
b) When not moving very far from the reference point, e.g., $\delta \leq 2/\nu$, then the overall
measurement cost is only by a small \emph{constant} factor more expensive than estimating a single
gradient vector; c) We generally prove that as we asymptotically approach the optimum,
analytic descent costs exactly the same as determining a single gradient vector.

%-------------------------
\begin{figure*}[tb]
	\begin{centering}
		\includegraphics[width=0.8\textwidth]{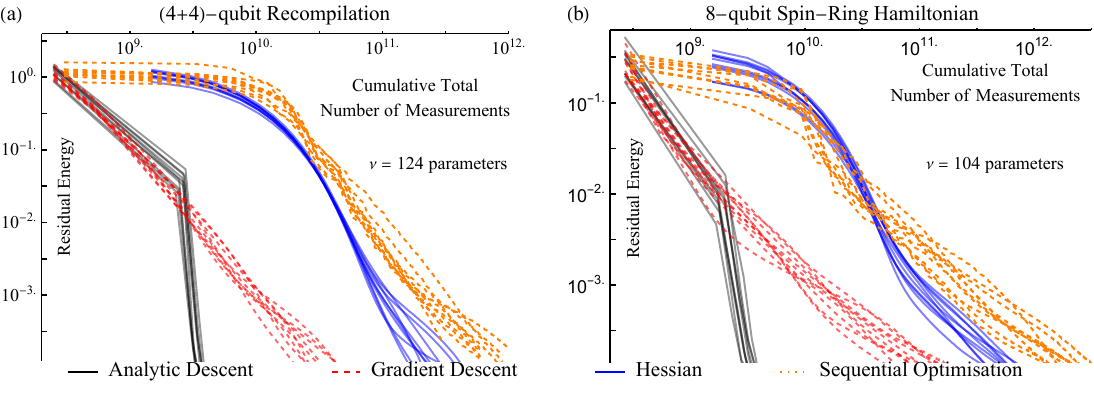}
		\vspace{-5mm}
		\caption{
				Distance from the exact ground-state energy (residual energy) as a function of the overall
				number of measurements (quantum resources).
				(a) Recompiling a 4-qubit unitary into hardware native gates via an $8$-qubit ground-state search problem
				and (b) finding the ground state of an $8$-qubit spin-ring Hamiltonian.
				Analytic descent appears to outperform all other techniques in terms of both convergence rate
				and the absolute level of quantum resources:
				its qualitative difference can be attributed to its ability 
				of explicitly keeping track of the evolution via an efficient classical approximation of the energy surface.
				In particular, a classical approximation of the energy surface is determined at each iteration step of analytic descent (solid lines) and in an internal loop we descent towards its minimum using a classical computer using
				gradient descent (not shown here).
				Our approximation is occasionally refined with optimally distributed additional measurements
				to keep shot noise (via $\epsilon^2$) below a threshold.
				Note that the hyperparameters have been optimised, especially the sampling rates,
				for each technique specifically so that the low energy regime can be reached.
				Consequently they do oversample in the early evolution and a left-to-right shift should be viewed as
				an artefact of this choice. All four techniques rely on determining the coefficients,
				such as $E^{(B)}_k$, from Eq.~\eqref{full-energy}.
				\label{fig2}
		}
	\end{centering}
\end{figure*}
%-------------------------

\section{Quantum Analytic Descent}
Instead of determining the gradient at every step, we use our classical approximation
of the \emph{entire} objective function $E(\underline{\theta})$ and its gradient vector
to descend towards its minimum using a classical computer. 
We propose an iterative optimisation in two nested loops.
First, in an external loop we use the quantum device to
estimate the coefficients in Eq.~\eqref{full-energy}
which allow us to build a classical
model of the full objective function around the reference point $\underline{\theta}_{0}$.
This initialisation costs exactly the same number of measurements as determining
a single gradient vector. 
In the internal loop, we compute our classical approximation of the gradient vector
at every iteration step and propagate our parameters $\underline{\theta}$
according to a suitable update rule.
With our efficient C code for computing the gradient vector,
descending 1000 steps towards the minimum can be performed in a
matter of minutes on a single thread for up to many hundreds of parameters \cite{gitcode}.

The internal, classical optimisation loop is aided with feedback from the
quantum device:
As we move away from the reference point,
shot noise in our classical approximation is magnified
via Eq.~\eqref{grad_variance_main} which would degrade the precision of our classical gradient.
Therefore, our optimal measurement distribution algorithm determines to which coefficients we
need to assign further measurements in order to keep this precision $\epsilon$ below
a threshold. For example,
when moving along a single slice $\theta_1$, then we need to use the quantum 
computer to sample only a \emph{linear number} of coefficients, namely the $E^{(D)}_{1,l}$.
Furthermore, note that our upper bound in Eq.~\eqref{meas_upb} 
guarantees that the overall number of additional measurements is
generally
proportional to the distance $\delta$ from the reference point.

Besides keeping the precision $\epsilon$ below a threshold, we also need to
ensure that our analytical approximation is valid (as it breaks down
for large $\delta$).
For example, one could
estimate the energy with the quantum device, e.g., at every $t$ iterations;
If the deviation from the analytical energy is too large
then the internal loop should abort and our approximation
in Eq.~\eqref{full-energy} should be re-initialised in the new reference point.
Other possibilities include, e.g., estimating the previously discussed similarity
measure $1-f$ or simply aborting when 
$\lVert \underline{\theta} \rVert_\infty$ or the iteration depth exceeds a certain threshold.

\section{Numerical simulations}
Let us now demonstrate our approach on
two problems of practical relevance.
We consider a hardware-efficient ansatz construction which is built of
alternating layers of parametrised single-qubit $X$ and $Y$ rotations and two-qubit
parametrised Pauli $ZZ$ gates as illustrated in Fig.~\ref{figansatz},
and we demonstrate our approach on two problems.
First, we consider an $8$-qubit
recompilation problem of a unitary operator $U$ that acts on $4$ qubits as in \cite{khatri2019quantumassisted}: since
recompiled unitaries may be repeated as part of a quantum
algorithm, they need to be determined to a very good approximation. Second, we
simulate a spin-ring Hamiltonian \cite{nandkishore2015many,childs2018toward} that is important in the context
of many-body localisation and we aim to determine its ground state to a high precision.

Fig.~\ref{fig2} shows the decreasing distance from the exact ground-state energy:
we do not compare the number of iterations but instead the number measurements (quantum resources).
As such, in Fig.~\ref{fig2}(black) analytic descent reaches the optimum faster than other techniques
(with fewer measurements and within about $5$ iterations).
Furthermore, analytic descent appears to have a considerably
accelerated asymptotic convergence rate, i.e., a significantly steeper slope,
even though 
in the early stages of the optimisation its efficacy is comparable
to simple gradient descent (red dashed lines).
The explantation is straightforward: building a classical approximation of a local region
may not be beneficial as we take big jumps in the early evolution.
In contrast, as we approach the optimum, our analytical approximation acts as a
	``memory'' and we can thus descend towards the optimum with very little overhead
	in measurement costs which is confirmed by the steep decrease in Fig~\ref{fig2}.
	Note also that the overall cost of  any optimisation algorithm is dominated
	by these later stages of evolution due to the fundamental shot-noise limit.

We show in the Appendix~\ref{hessianSection} that the Hessian is determined by our coefficients in
Eq.~\eqref{full-energy} and similarly provides an $\mathcal{O}(\delta^3)$ approximation
of the local energy surface. Fig.~\ref{fig2}(blue) confirms
that initially the Hessian-based Newton-Raphson approach is faster than gradient descent (steeper slope),
but then slows down when approaching the optimum for three main reasons
as expected from ref.~\cite{van2021measurement}. a) As opposed to analytic descent,
we need to determine all the $\mathcal{O}(\nu^2)$ coefficients to a relatively high precision
for computing the inverse of the ill-conditioned Hessian;
b) the measurement cost grows with $\eta^4$ of a regularisation parameter $\eta$, and we therefore
set $\eta = 0.1$ to keep costs practical -- whereas an increased $\eta$
reduces convergence rate;
c) We use the Hessian to determine a jump in parameter space, but this jump is taken with respect
to a Euclidean geometry as opposed to the relevant Riemannian geometry with substantial off-diagonal
entries in the metric tensor \cite{koczor2019quantum,quantumnatgrad}.

Fig.~\ref{fig2}(dashed orange) shows the sequential optimisation approach from refs.~\cite{PhysRevResearch.2.043158,parrish2019jacobi,ostaszewski2019quantum},
whereby we repeatedly jump to the global minimum along single parameter slices $\theta_k$.
The approach is initially faster than gradient descent (steeper slope).
We note that hyperparameters, in particular the sampling rate, of each technique have been specifically optimised such
that a convergence criterion $\Delta E = 10^{-4}$ can be reached. We therefore inevitably oversample in the early evolution
and a left-to-right shift in Fig.~\ref{fig2} should be viewed as an artefact: in a low-precision setting,
e.g., $\Delta E = 10^{-2}$, sequential optimisation may even outperform others \cite{PhysRevResearch.2.043158,parrish2019jacobi,ostaszewski2019quantum}.

We finally remark that quantum natural gradient has
been shown in numerical studies to significantly outperform classical optimisers and to be less
vulnerable to getting stuck in local optima \cite{Li2017,samimagtime,koczor2019quantum,xiaotheory,quantumnatgrad,wierichs2020avoiding}.
We numerically demonstrate in Appendix~\ref{natgrad} that analytic descent is further enhanced by taking into account the
metric information by building a classical approximation of the quantum Fisher information $[\qf]_{mn}$.
In particular, the metric tensor entries can also be approximated classically as
\begin{equation}\label{metric-tensor-eq}
	[\qf]_{mn} =
	\mathcal{F}_{BB} F_{BB}(\underline{\theta})
	+ \mathcal{F}_{AB} F_{AB}(\underline{\theta}) + \dots  \mathcal{O}(\sin^2\delta),
\end{equation} 
where $\mathcal{F}_{BB}$ are real coefficients that we can estimate by computing overlaps
between quantum states while $F_{BB}(\underline{\theta})$ are trigonometric monomials.

\section{Conclusion and Discussion}

In this work we considered analytical characterisations
of variational quantum circuits that are composed of Pauli gates; although exponentially many coefficients determine
a full trigonometric expansion, we propose an efficient, approximate approach
for characterising the ansatz landscape in the vicinity of any reference point.

We propose a novel optimisation technique: a quantum device is used to determine
a classical approximation of the entire energy surface. 
A classical optimisation routine is then used in an internal loop to descend towards the minimum of this approximate surface.
We have devised an exact, optimal measurement distribution strategy whereby
the quantum computer is occasionally used to perform further targeted measurements
to reduce shot noise in our classical model: we generally prove that asymptomatically
the measurement cost of an entire `jump’ in our approach corresponds to determining just a single gradient vector.

We numerically simulated practical problems and
observed that indeed analytic descent significantly outperforms other techniques
both in terms of the number of measurements and in terms of its convergence rate.
We have made our efficient C implementation of the approach publicly available \cite{gitcode}.

There are a number of apparent, promising extensions. First, we could use the
information from the previous iterations as a Bayesian prior when re-estimating
our classical model in a next step. Second, we can similarly build a classical
model of the quantum Fisher information matrix and compute it in the internal optimisation classically
without using the quantum device.

We note that our approach is completely general and can
be applied to any Hamiltonain $\mathcal{H}$, although, we expect that increasingly more
complex Hamiltonians -- such as in quantum chemistry -- might result in more complex
energy surfaces which are more difficult to approximate classically.
Nevertheless, a significant advantage of our approach is that in all cases it guarantees an approximation error
of the gradient vector
that scales with the fourth power of the distance from the reference point
as shown in Fig.~\ref{fig1}(b).
While our analytical approximation may be accurate for
	relatively large jumps $\delta$, we have shown that its measurement cost relative to
	determining a gradient vector grows with the distance $\delta$.

	As such, the main limitation of the present approach is that in the early evolution
	it may be less beneficial to build a local approximation of the energy surface
	due to the increased sampling costs.
	The present work therefore motivates a hybrid approach whereby analytic descent
	complements other techniques:
	in the early evolutions one may benefit from, e.g., applying a sequential
	optimisation~\cite{PhysRevResearch.2.043158,parrish2019jacobi,ostaszewski2019quantum}
	or natural gradient~\cite{samimagtime,koczor2019quantum,xiaotheory,quantumnatgrad,wierichs2020avoiding},
	while in the later stages of the evolution one would switch to analytic descent. 
However, it is important to recognise that the bulk of the optimisation costs are absorbed by
	the later stages of the evolution.
For example, in Fig.~\ref{fig2} we spend less than $10^{9.5}$ shots in the early stages while
it takes an order of magnitude more, $10^{10.5}$ shots, to reach our convergence criterion with standard
gradient descent. As such, Quantum Analytic Descent could reduce the overall cost of optimisation by at least
an order of magnitude, and this figure is further increased when using more advanced techniques
for adaptively setting sampling rates.

\begin{acknowledgments}
SCB acknowledges financial support from EPSRC Hub grants under the agreement numbers
EP/M013243/1 and EP/T001062/1, and from the IARPA funded LogiQ project.
BK and SCB acknowledge funding received from EU H2020-FETFLAG-03-2018 under the grant
agreement No 820495 (AQTION). 
BK thanks the University of Oxford for a Glasstone Research Fellowship and Lady Margaret Hall, Oxford for a Research Fellowship.
The numerical modelling involved in this study made use of the Quantum Exact Simulation Toolkit (QuEST), and the recent development QuESTlink\,\cite{QuESTlink} which permits the user to use Mathematica as the integrated front end. The authors are grateful to those who have contributed to both these valuable tools. 
The views and conclusions contained herein are those of
the authors and should not be interpreted as necessarily representing the official policies or endorsements, either expressed or implied, of the ODNI, IARPA, or the
U.S. Government. The U.S. Government is authorized to
reproduce and distribute reprints for Governmental purposes notwithstanding any copyright annotation thereon.
Any opinions, findings, and conclusions or recommendations expressed in this material are those of the author(s) and do not necessarily reflect the view of the
U.S. Army Research Office. 
Let us finally remark that the present technique has recently been extended to general quantum gates in ref.~\cite{genpshift}.
\end{acknowledgments}

\appendix

\section{Quantum gates generated by Pauli strings}
\subsection{Expressing a single gate \label{singlePauliSec}}
Let us consider a single gate in the ansatz circuit $U_k(\theta_k) $, where $k$
indexes its position and $k\in \{1, 2, \dots \nu\}$ with $\nu$ denoting the number of parameters. We assume that this gate
is generated by a Pauli string $P_k$ and ideally (when the gate is not noisy), it corresponds to
the following unitary operator
\begin{align}
	U_k(\theta_k) :=& \exp(- i \theta_k P_k/2)\\
	 =& \cos[\theta_k/2] \mathrm{Id} - i \sin[\theta_k/2] P_k,
\end{align}
where the second equality straightforwardly follows from the algebra $P_k^{2n}=\mathrm{Id}$ and $P_k^{2n+1}=P_k$.

We now fix the parameter dependence of this gate at the reference point $\theta_0$ and express the
action of this gate on any quantum state using the continuous angle $\theta$.
Let us first define the quantum gate as a mapping $\Phi_k(\theta): \mathcal{D} \mapsto \mathcal{D}$
over density operators, where $\mathcal{D}$ denotes the set of density operators, i.e.,
positive, unit trace operators over the Hilbert space $\mathbb{C}^{2^N}$.
The gate can then be expressed as a general mapping over arbitrary density matrices $\rho$
as the unitary conjugation $U_k(\theta_0+\theta) \rho U_k^\dagger(\theta_0+\theta)$, and this can be expanded
into the following	transformations
\begin{align} \label{signlegate-derivation}
	\Phi_k(\theta) \rho := & U_k(\theta) U_k(\theta_0) \rho U_k^\dagger(\theta_0) U_k^\dagger(\theta)\\\nonumber
	 = &
	\cos^2[\theta_k/2] \rho_{ref} + \sin^2[\theta_k/2]  P_k \rho_{ref} P_k  \\\nonumber
		& - i \cos[\theta/2]\sin[\theta/2] (P_k \rho_{ref} - \rho_{ref} P_k ).
\end{align}
Here we have used the notation $\rho_{ref} := U_k(\theta_0) \rho U_k^\dagger(\theta_0)$.
The dependency on the continuous angle $\theta$ is absorbed into the following functions
\begin{align*}
	& \cos[\theta/2]^2  = (1+\cos[\theta])/2, \\
	& \cos[\theta/2]\sin[\theta/2] = \sin[\theta]/2, \\
	& \sin[\theta/2]^2  = (1-\cos[\theta])/2 .
\end{align*}

We can now formalise Eq.~\eqref{signlegate-derivation} by separating it into
\emph{discrete} mappings over density matrices which are multiplied by continuous functions 
that depend on the parameter $\theta$ as
\begin{equation}\label{single-gate-expression}
	\Phi_k(\theta) = 
	a(\theta) \Phi_{ak}
	+ b(\theta) \Phi_{bk}
	+ c(\theta) \Phi_{ck}.
\end{equation}
Here the mapping depends on the
parameter $\theta$ via the Fourier components $a(\theta),b(\theta),c(\theta) : \mathbb{R} \mapsto \mathbb{R}$
and we define their explicit forms as 
\begin{align} \label{single-functions}
	a(\theta) := &  (1+\cos[\theta])/2 = \mathcal{O}(1+\theta^2),\\
	b(\theta) := 	& \sin[\theta]/2  = \mathcal{O}(\theta),\\
	c(\theta) := 	&  (1-\cos[\theta])/2 = \mathcal{O}(\theta^2),
\end{align}
and we have also included their scaling when approaching $\theta \rightarrow 0$.
Note that  we have intentionally introduced the constant shift $\theta_0$ and, of course, our definition
corresponds to the action $\Phi_k(0) [\rho] = U_k(\theta_0) \rho U_k^\dagger(\theta_0)$
for the case $\theta \rightarrow 0$.
The discrete mappings $\Phi_{ak},\Phi_{bk}$ and $\Phi_{ck}$ in Eq.~\eqref{single-gate-expression}
can be specified via their action on arbitrary density matrices as
\begin{align*}
	\Phi_{ak} \, \rho =& U_k(\theta_0) \rho U_k^\dagger(\theta_0) \equiv \Phi_k(0) \rho,\\[2mm]
	\Phi_{bk} \, \rho =& - i [P_k,U_k(\theta_0) \rho U_k^\dagger(\theta_0)]\\
	 =&
	-\frac{\partial \left( U_k(\theta) U_k(\theta_0) \rho U_k^\dagger(\theta_0) U_k^\dagger(\theta) \right)}{\partial \theta} |_{\theta = 0}\\
	=&  U_+\rho U_+^\dagger -  U_- \rho U_-^\dagger \equiv [\Phi_k(\pi/2) - \Phi_k(-\pi/2)]\rho,	\\[2mm]
	\Phi_{ck} \, \rho =& P_k \, U_k(\theta_0) \rho U_k^\dagger(\theta_0) \, P_k^\dagger =  U_k(\theta_0+\pi) \rho U_k^\dagger(\theta_0+\pi)\\
	\equiv&	 \Phi_k(\pi) \rho,
\end{align*}
where we have  denoted $U_+:=U_k(\theta_0{+}\pi/2)$ and $U_-:=U_k(\theta_0{-}\pi/2)$.
We finally conclude by recollecting their explicit forms as
\begin{align}\label{single-mappings}
	\Phi_{ak} &= \Phi_k(0),   \quad \quad		\Phi_{bk} = \Phi_k(\pi/2) - \Phi_k(-\pi/2), \\\nonumber
	\Phi_{ck}  &=\Phi_k(\pi).
\end{align}

We can use the above expressions to express any linear mapping, such as the energy functional
$ \mathcal{E}(\rho) : \mathcal{D} \mapsto \mathbb{R} $,	 via the 
trace relation	$\mathcal{E}(\rho) = \tr[\mathcal{H}^\dagger \rho]$,
which is often referred to as an expectation value,
and $\mathcal{H}$ is any Hermitian operator in the Hilbert space $\mathbb{C}^{2^N}$.
We now consider the parametric mapping $E(\theta) :  \mathbb{R} \mapsto \mathbb{R}$,
which we define as $E(\theta) := [\mathcal{E} \circ \Phi_k(\theta) ]\rho_0= \mathcal{E}(\Phi_k(\theta) \rho_0)$
and we refer to it as the energy function, or energy landscape. The reference state
can be, e.g., the computational zero state $\rho_0 := |\underline{0}\rangle \langle \underline{0} |$.
We can express the energy function explicitly via the
following Fourier series
\begin{equation}
	E(\rho) = \tr[\mathcal{H} \, \Phi_k(\theta) \rho_0 ]
	=  \alpha_k a(\theta)  +  \beta_k b(\theta)  +  \gamma_k c(\theta).
\end{equation}
The Fourier coefficients $\alpha_k, \beta_k, \gamma_k \in \mathbb{R}$ can be completely determined by discrete
samples of the energy function via the discrete mappings of the density matrix as
\begin{align}
	\alpha_k :=& \tr[\mathcal{H} \, \Phi_{ak} \rho_0 ] = E(0) + E(\pi)\\
	\beta_k :=& \tr[\mathcal{H} \, \Phi_{bk}\rho_0 ] = E(\pi)\\
	\gamma_k :=& \tr[\mathcal{H} \, \Phi_{ck} \rho_0 ] = E(\pi/2) - E(-\pi/2).
\end{align}

The above formula informs us that we can completely and analytically determine the full energy function
$E(\theta)$,  just by 
querying the function $E(\theta)$ at four different points as $(-\pi/2, 0, \pi/2, \pi)$. Of course
Nyquist's theorem also informs us that this is suboptimal, since the Fourier spectrum of $E(\theta)$ is
bounded with only 3 frequency terms present $(-1,0,1)$. This guarantees that querying the function $E(\theta)$
at only 3 points would be sufficient for completely reconstructing it.
Note that due to our definitions, the parameter $\theta$ is shifted by the constant $\theta_0$
and, for example, $E(0) = \tr[ \mathcal{H} U_k(\theta_0) \rho_0 U_k^\dagger(\theta_0)]$.

\subsection{Expanding the full ansatz circuit \label{fullansatz-section}}

Let us now consider the effect of the full ansatz circuit
on the reference state $\rho_0 := |\underline{0}\rangle \langle \underline{0} |$ as 
$U(\underline{\theta}_0+\underline{\theta}) \rho_0  U^\dagger(\underline{\theta}_0+\underline{\theta})$
with using the notation
\begin{equation*}
U(\underline{\theta}_0 {+} \underline{\theta}) := U_\nu(\theta_{0,\nu} {+}\theta_\nu) \cdots U_2(\theta_{0,2}{+}\theta_2) U_1(\theta_{0,1} {+} \theta_1).
\end{equation*}
Here $\underline{\theta}_0 \in \mathbb{R}^\nu$ is a vector that represents a fixed, constant shift of the parameters, while
the circuit depends continuously on the parameters $\underline{\theta}\in \mathbb{R}^\nu$.

Using results from the previous subsection, we can build an analytical model of the superoperator representation
$\Phi(\underline{\theta})$ 	of the full ansatz circuit as the mapping 
\begin{align}
	\Phi(\underline{\theta}):=& \Phi_\nu(\theta_\nu) \dots \Phi_2(\theta_2) \Phi_1(\theta_1) \\\nonumber
	= &
	\prod_{k=1}^\nu
	[a(\theta_k) \Phi_{ak}
	+ b(\theta_k) \Phi_{bk}
	+ c(\theta_k) \Phi_{ck}].
\end{align}	
The above equation expresses the full ansatz circuit and its dependence on the parameters $\underline{\theta}$.
Of course fully expanding the above expression would result in a sum of $3^\nu$ different terms. Nevertheless, 
we expand this into a sum and truncate the expansion such that the remaining terms are correct up to an error $
\mathcal{O}(\sin^3\delta)$. For this we define $\delta := \lVert \underline{\theta}\rVert_\infty$, to denote the
absolute largest entry in the vector $\underline{\theta}$.
We assume that the continuous parameters are only used to
explore the vicinity of the reference point $\underline{\theta}_0$ in parameter space
via a sufficiently small $\delta$.
This can be, e.g., when the reference parameters $\underline{\theta}_0$ are already a good approximation of the optimal ones
as $\lVert \underline{\theta}_0 - \underline{\theta}_{opt} \rVert_\infty < \delta$ with $\delta \ll 1$ and we search for the
ground state energy by optimising the continuous parameters.

Let us now derive our approximation: we first substitute the explicit forms of
the trigonometric functions into the expression above as
\begin{equation*}
	\prod_{k=1}^\nu
		[\frac{1{+}\cos(\theta_k)}{2} \Phi_{ak}
		{+} \frac{\sin(\theta_k)}{2} \Phi_{bk}
		{+} \frac{1{-}\cos(\theta_k)}{2} \Phi_{ck}],
\end{equation*}
and expand this product into a sum of $3^\nu$ terms.
We drop all terms that have a product of 3 or more $\sin(\theta_k)$ terms in them 
thereby obtaining an approximate mapping that is correct up to $\mathcal{O}(\sin^3\delta)$
as
\begin{align} \label{approx-mapping}	
	\tilde{\Phi}(\underline{\theta}) :=& A(\underline{\theta}) \Phi^{(A)} + \sum_{k=1}^\nu [B_k(\underline{\theta}) \Phi^{(B)}_k + C_k(\underline{\theta}) \Phi^{(C)}_k]\\\nonumber
	 &+
	\sum_{k}^\nu\sum_{l=k+1}^\nu [D_{kl}(\underline{\theta}) \Phi^{(D)}_{kl}].
\end{align}
Here the functions $A(\underline{\theta})$, $B_k(\underline{\theta})$, $C_k(\underline{\theta})$ and $D_{kl}(\underline{\theta})$ absorb the dependence
on the parameters $\underline{\theta}$ and $\Phi^{(A)}_k$, $\Phi^{(B)}_k$, $\Phi^{(C)}_k$ and $\Phi^{(D)}_{kl}$ are superoperators of discrete mappings.
We compute the explicit form of the terms appearing in the summation in Eq.~\eqref{approx-mapping} as
\begin{align*}
	A(\underline{\theta})\times \Phi^{(A)}  =& \prod_{k=1}^\nu [a(\theta_k) \Phi_{ak}] = \mathcal{O}(1), \\
	B_k(\underline{\theta}) \times \Phi^{(B)}_k =&   a(\theta_\nu) a(\theta_{\nu-1}) \cdots b(\theta_k)  \cdots a(\theta_2) a(\theta_1) \\
	&\times \Phi_{a\nu} \Phi_{a(\nu-1)} \cdots \Phi_{bk} \cdots  \Phi_{a2} \Phi_{a1}\\
	 =& \mathcal{O}(\theta_k),\\
	C_k(\underline{\theta}) \times \Phi^{(C)}_k =&   a(\theta_\nu) a(\theta_{\nu-1}) \cdots c(\theta_k) \cdots  a(\theta_2) a(\theta_1)\\
	&\times  \Phi_{a\nu}  \Phi_{a(\nu-1)} \cdots  \Phi_{ck} \cdots \Phi_{a2} \Phi_{a1}\\
	 =& \mathcal{O}(\theta_k^2),	\\
	D_{kl}(\underline{\theta}) \times \Phi^{(D)}_{kl} =& 
	a(\theta_\nu) a(\theta_{\nu-1}) \cdots b(\theta_k) \cdots b(\theta_l) \cdots   a(\theta_1) \\
	& \times
	\Phi_{a\nu} \Phi_{a(\nu-1)} \cdots \Phi_{ck} \cdots \Phi_{cl} \cdots \Phi_{a1}\\
	 =& \mathcal{O}(\theta_k \theta_l ).
\end{align*}
The discrete mappings can be further simplified by using Eq.~\eqref{single-mappings} as
\begin{align}
	\Phi^{(A)} =& \Phi(\underline{0}), \\\nonumber
	\Phi^{(B)}_k =& \Phi(\tfrac{1}{2}\pi \underline{v}_k) - \Phi(-\tfrac{1}{2}\pi \underline{v}_k),\\\nonumber
	\Phi^{(C)}_k =& \Phi(\pi \underline{v}_k)\\\nonumber
	 \Phi^{(D)}_{kl} =& 
	\Phi(\tfrac{1}{2}\pi \underline{v}_k + \tfrac{1}{2}\pi \underline{v}_l )
	+\Phi(-\tfrac{1}{2}\pi \underline{v}_k - \tfrac{1}{2}\pi \underline{v}_l )\\\nonumber
	&-\Phi(-\tfrac{1}{2}\pi \underline{v}_k + \tfrac{1}{2}\pi \underline{v}_l )
	-\Phi(\tfrac{1}{2}\pi \underline{v}_k - \tfrac{1}{2}\pi \underline{v}_l ),
\end{align}
where $\underline{v}_k\in \mathbb{R}^\nu$ denotes the standard basis vector, e.g.,  $(0,0,\dots 0,1, 0, \dots 0 )$.
We further remark that due to our definitions, the parameters $\underline{\theta}$ are shifted by the constant vector $\underline{\theta}_0$
and, for example, $\Phi(\underline{0})\rho_0 =  U(\underline{\theta}_0) \rho_0 U^\dagger(\underline{\theta}_0)$.

We can quantify the error of the approximate mapping in Eq.~\eqref{approx-mapping} via the 
trace distance of the resulting density operators and we express this as
$\lVert \Phi(\underline{\theta}) \rho - \tilde{\Phi}(\underline{\theta}) \rho \rVert_{tr} =\mathcal{O}(\sin^3\delta) $.
We remark that our expansion in Eq.~\eqref{approx-mapping}
consist of a sum of $1+\nu+ \nu^2/2$ different terms and describes the variational mapping up to an error
$\mathcal{O}(\sin^3\delta)$. 
We could similarly expand the mapping into a sum of $\mathcal{O}(\nu^3)$ terms and have an error $\mathcal{O}(\sin^4\delta)$
or beyond.
	As such, in general we can obtain a family of approximations to the energy landscape:
	by discarding all terms that contain a product of $q$ or more $\sin\theta_k$ terms we obtain
	an approximation as a sum of $\mathcal{O}(\nu^{q-1})$ terms with an approximation error $\mathcal{O}(\sin^q \delta)$.

\section{Approximating the full energy surface locally \label{energySection}}
We can express the full energy surface following our definition in the previous section 
and evaluating the discrete mappings
\begin{align}\label{surfaceapprox}
	E(\underline{\theta}) :=& \tr[\mathcal{H} \,  \Phi(\underline{\theta})\rho_0]
	=A(\underline{\theta})  E^{(A)} \\\nonumber
	 &+ \sum_{k=1}^\nu [B_k(\underline{\theta}) E^{(B)}_k
	+ C_k(\underline{\theta}) E^{(C)}_k]\\\nonumber
	&+
	\sum_{k}^\nu\sum_{l=k+1}^\nu [D_{kl}(\underline{\theta}) E^{(D)}_{kl}] + \mathcal{O}(\sin^3 \delta).
\end{align}
We can express the discrete mappings as queries of the energy function at
discrete points in parameter space as
\begin{align*} \label{energymeasurments}
	E^{(A)} =& \tr[ \mathcal{H} \, \Phi^{(A)} \rho_0 ] = E(\underline{0})\\
	E^{(B)}_k =&  \tr[\mathcal{H} \, \Phi^{(B)}_k \rho_0 ]
	=  E(\tfrac{1}{2}\pi \underline{v}_k) -   E(-\tfrac{1}{2}\pi \underline{v}_k)\\
	E^{(C)}_k =&  \tr[ \mathcal{H} \, \Phi^{(C)}_k \rho_0 ] =  E(\pi \underline{v}_k) \\
	E^{(D)}_{kl} =&  \tr[ \mathcal{H} \,  \Phi^{(D)}_{kl}  \rho_0 ] =\\
	&E(\tfrac{1}{2}\pi \underline{v}_k {+} \tfrac{1}{2}\pi \underline{v}_l )
	+  E(-\tfrac{1}{2}\pi \underline{v}_k {-} \tfrac{1}{2}\pi \underline{v}_l )\\
	&-	  E(-\tfrac{1}{2}\pi \underline{v}_k {+} \tfrac{1}{2}\pi \underline{v}_l )
	-  E(\tfrac{1}{2}\pi \underline{v}_k {-} \tfrac{1}{2}\pi \underline{v}_l ).
\end{align*}
Here $\underline{v}_k\in \mathbb{R}^\nu$ denotes a standard basis vector, e.g.,  $(0,0,\dots 0,1, 0, \dots 0 )$.
Note that due to our definitions, the parameters $\underline{\theta}$ are shifted by the constant vector $\underline{\theta}_0$
and, for example, $E(\underline{0}) = \tr[ \mathcal{H} U(\underline{\theta}_0) \rho_0 U^\dagger(\underline{\theta}_0)]$.

Using the above expressions, one can determine an $\mathcal{O}(\sin^3\delta)$
approximation of the full energy surface by querying the energy function $E(\underline{\theta}) $
at a total number of $Q$ points, where
\begin{equation}
	Q = 1 + \nu + 2 \nu + 4(\nu^2/2-\nu) = 1 + 2\nu^2 -2 \nu.
\end{equation}

\subsection{Expressing the gradient analytically \label{gradientSection}}
We now derive the dependence of the gradient vector components $g_m := \partial_m E(\underline{\theta})$ on
the parameters $\underline{\theta}$ using our approximation from Eq.~\eqref{surfaceapprox}.
We can explicitly write
\begin{align}\nonumber
	\partial_m E(\underline{\theta})
	=& \frac{\partial A(\underline{\theta}) }{\partial \theta_m} E^{(A)}   {+} \sum_{k=1}^\nu [ \frac{\partial  B_k(\underline{\theta})}{\partial \theta_m} E^{(B)}_k {+}
	\frac{ \partial C_k(\underline{\theta}) }{\partial \theta_m} E^{(C)}_k]\\
	\label{full-grad-equation}
	&+
	\sum_{k}^\nu\sum_{l=k+1}^\nu [\frac{ \partial D_{kl}(\underline{\theta})}{\partial \theta_m} E^{(D)}_{kl}] + \mathcal{O}(\sin^2\delta).
\end{align}
Let us fist compute the derivatives of the single-variate functions from Eq.~\eqref{single-functions} as
\begin{align*}
	\frac{\partial a(\theta_k)}{\partial \theta_k}&= - \sin[\theta_k]/2, \quad \quad
	\frac{\partial b(\theta_k)}{\partial \theta_k}=  \cos[\theta_k]/2, \\
	\frac{\partial c(\theta_k)}{\partial \theta_k}&=  \sin[\theta_k]/2.
\end{align*}
We now compute partial derivatives of the monomials; The first term is
\begin{equation*}
	\frac{\partial A(\underline{\theta}) }{\partial \theta_m}  = a(\theta_\nu) a(\theta_{\nu-1}) \cdots \frac{\partial a(\theta_m)}{\partial \theta_m}  \cdots  a(\theta_1) = \mathcal{O}(\theta_m).
\end{equation*}
\begin{widetext}
Similarly we have
\begin{equation}
	\frac{\partial B_k(\underline{\theta})  }{\partial \theta_m} =
	\begin{cases} 
		a(\theta_\nu) a(\theta_{\nu-1}) \cdots \frac{\partial b(\theta_m)}{\partial \theta_m}  \cdots a(\theta_2) a(\theta_1)  =  \mathcal{O}(1)
		& \mbox{if } 	k = m\\
		a(\theta_\nu) a(\theta_{\nu-1}) \cdots b(\theta_k)  \cdots \frac{\partial a(\theta_m)}{\partial \theta_m}  \cdots a(\theta_2) a(\theta_1) 
		= \mathcal{O}(\theta_m \theta_k)
		& \mbox{if } 	k \neq m,\\
	\end{cases}
\end{equation}
but note that here we do not not assume that $m>k$.
Very similarly we have
\begin{equation}
	\frac{\partial C_k(\underline{\theta})  }{\partial \theta_m} =
	\begin{cases} 
		a(\theta_\nu) a(\theta_{\nu-1}) \cdots \frac{\partial c(\theta_m)}{\partial \theta_m}  \cdots a(\theta_2) a(\theta_1)  =  \mathcal{O}(\theta_m)
		& \mbox{if } 	k = m\\
		a(\theta_\nu) a(\theta_{\nu-1}) \cdots c(\theta_k)  \cdots \frac{\partial a(\theta_m)}{\partial \theta_m}  \cdots a(\theta_2) a(\theta_1)
		= \mathcal{O}(\theta_m \theta_k^2	)
		& \mbox{if } 	k \neq m.\\
	\end{cases}
\end{equation}
Finally,
\begin{equation}
	\frac{ \partial D_{kl}(\underline{\theta})}{\partial \theta_m} = 
	\begin{cases} 
		a(\theta_\nu) a(\theta_{\nu-1}) \cdots \frac{\partial b(\theta_m)}{\partial \theta_m}  \cdots b(\theta_l) \cdots a(\theta_2) a(\theta_1)  =  \mathcal{O}(\theta_m)
		& \mbox{if } 	k = m\\
		a(\theta_\nu) a(\theta_{\nu-1}) \cdots b(\theta_k) \cdots \frac{\partial b(\theta_m)}{\partial \theta_m}   \cdots a(\theta_2) a(\theta_1)  =  \mathcal{O}(\theta_m)
		& \mbox{if } 	l = m\\
		a(\theta_\nu) a(\theta_{\nu-1}) \cdots b(\theta_k)  \cdots b(\theta_l) \cdots \frac{\partial a(\theta_m)}{\partial \theta_m	}  \cdots a(\theta_2) a(\theta_1)  = \mathcal{O}(\theta_k \theta_l \theta_m )
		& \mbox{if } 	k \neq m \neq l.\\
	\end{cases}
\end{equation}

%-------------------------
\begin{figure*}[tb]
	\begin{centering}
		\includegraphics[width=0.45\textwidth]{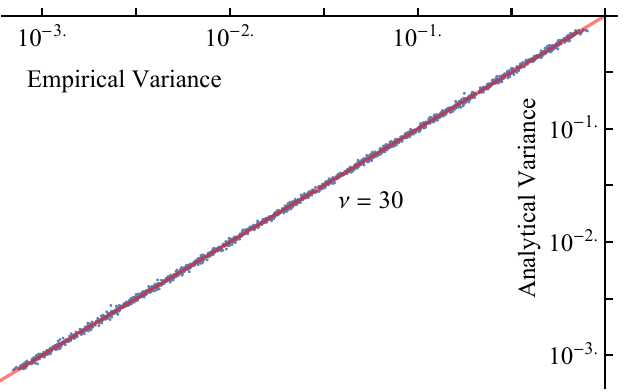}
		\includegraphics[width=0.45\textwidth]{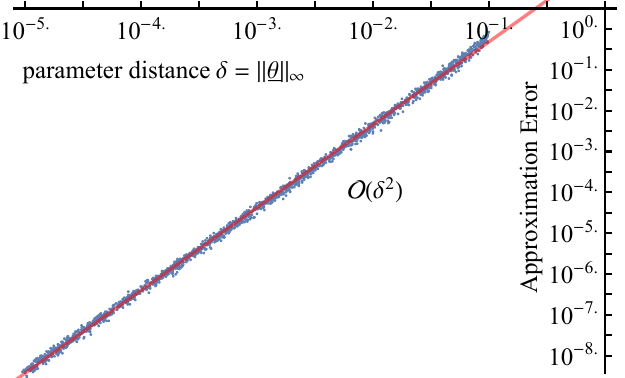}
		\caption{
			(left) Empirically estimating the precision $\epsilon^2$ (variance) as
			the expected Euclidean distance from the exact gradient vector $\epsilon^2:=\langle \lVert\Delta g\rVert^2 \rangle = \sum_{k=1}^\nu \var[ \partial_m E(\underline{\theta}) ]$
			for $2000$ randomly selected points in parameter space.
			This verifies our analytical expression derived in Sec.~\ref{sec:opt_measurement}
			that we have numerically exactly computed using our efficient C code \cite{gitcode}.
			(right) We compute the exact expression for the function $T(\underline{\theta})$ as defined in Eq.~\eqref{tdefinition}
			using our efficient C code and compare it to the analytical approximation in
			\eqref{tapproximation} and obtain the expected $\mathcal{O}(\delta^2)$ error term.
			The analytical approximation in \eqref{tapproximation}  is used to derive the
			scaling of the measurement cost of the analytic descent approach.
			\label{check_variances_tapprox}
		}
	\end{centering}
\end{figure*}
%-------------------------
\end{widetext}
One can therefore compute the full gradient vector analytically, up to an error $\mathcal{O}(\sin^2\delta)$,
via the monomials $A(\underline{\theta})$, $B_k(\underline{\theta})$, $C_k(\underline{\theta})$ and $D_{kl}(\underline{\theta})$
and the corresponding energy coefficients.
These coefficients can be determined by querying the energy function at $\mathcal{O}(\nu^2)$ points
as discussed in Sec.~\ref{energySection}.
We propose an efficient  classical algorithm for computing this gradient 
vector, and its computational complexity is $\mathcal{O}(\nu^3)$, refer to Sec.~\ref{classical-alg-sec}.

\subsection{Error propagation and variances\label{variance-sec}}
Using the usual linear error propagation formula, the variance of the gradient estimator can be computed 
via the following terms
\begin{align}\label{grad_var}
	&\var[ \partial_m E(\underline{\theta}) ]
	= [\frac{\partial A(\underline{\theta}) }{\partial \theta_m}]^2 \, \var[ E^{(A)}]   \\\nonumber
	&+ \sum_{k=1}^\nu 
	\Bigg(  [ \frac{\partial  B_k(\underline{\theta})}{\partial \theta_m}]^2 \, \var[ E^{(B)}_k] + [\frac{ \partial C_k(\underline{\theta}) }{\partial \theta_m}]^2 \, \var[ E^{(C)}_k]^2 \Bigg) \\\nonumber
	&+
	\sum_{k}^\nu\sum_{l=k+1}^\nu \Bigg(
	[\frac{ \partial D_{kl}(\underline{\theta})}{\partial \theta_m}]^2 \,  \var[E^{(D)}_{kl}]
	\Bigg) \nonumber
\end{align}
Here the variances, such as $\var[ E^{(A)}] = \var[ E(\underline{0})] $, are directly proportional to the
precision of the energy estimation. This variance scales inversely with how many times the energy estimator is 
sampled.

Now using the scaling of the multivariate functions from \ref{gradientSection}, we can expand the above variance 
into a leading term $[\frac{\partial  B_m(\underline{\theta})}{\partial \theta_m}]^2 = \mathcal{O}(1) $
and into terms that scale with $\delta$ as
\begin{equation*}
	\var[ \partial_m E(\underline{\theta}) ] =  [\frac{\partial  B_m(\underline{\theta})}{\partial \theta_m}]^2 \, \var[ E^{(B)}_k]
	+ \mathcal{O}(\sin^2\delta) .
\end{equation*}
As long as the norm $\lVert \underline{\theta} \rVert_\infty < \delta$ is sufficiently small,
the variance of the gradient vector is dominated by the variances of measuring $\var[ E^{(B)}_k]$.
This means that, even though one has to query the energy function at $\mathcal{O}(\nu^2)$
points, most of these queries need not be very precise. In fact, the variance
of the gradient component is dominated by the precision of the $\mathcal{O}(\nu)$
queries used to determine the coefficients 	$E^{(B)}_k$. Conversely, the 
measurement cost of estimating our classical model to a high precision is dominated by
estimating the coefficients $E^{(B)}_k$. Let us now derive an optimal measurement strategy that confirms these expectations.

\subsection{Optimal measurement distribution \label{sec:opt_measurement}}	

Using techniques from \cite{van2021measurement} we now derive an optimal measurement strategy for estimating coefficients 
in our analytical approximation of the gradient vector, refer also to~\cite{rubin2018application}.
Let us define the precision of determining the full gradient vector via the expected euclidean distance from the
mean as $\epsilon^2:= \langle \lVert\Delta g\rVert^2 \rangle = \sum_{m=1}^\nu \var[ \partial_m E(\underline{\theta}) ]$. We can express this precision
explicitly using the above variance propagation formula as
\begin{align*}
	\epsilon^2 
	=&   \sum_{m=1}^\nu [\frac{\partial A(\underline{\theta}) }{\partial \theta_m}]^2  \, \var[ E^{(A)}]  +
	\sum_{m,k=1}^\nu [ \frac{\partial  B_k(\underline{\theta})}{\partial \theta_m}]^2 \, \var[ E^{(B)}_k]\\
	&+ \sum_{m,k=1}^\nu [\frac{ \partial C_k(\underline{\theta}) }{\partial \theta_m}]^2 \, \var[ E^{(C)}_k]  \\
	&+
	\sum_{k=1}^\nu\sum_{l=k+1}^\nu \Bigg(
	\sum_{m=1}^\nu [\frac{ \partial D_{kl}(\underline{\theta})}{\partial \theta_m}]^2 \,  \var[E^{(D)}_{kl}]
	\Bigg).
\end{align*}
Let us simplify the above equation by introducing the abbreviations that denote the following trigonometric
polynomials as
\begin{align} \label{calCoeffs}
	\mathcal{A} &:= \sum_{m=1}^\nu [\frac{\partial A(\underline{\theta}) }{\partial \theta_m}]^2,
	\quad \quad
	\mathcal{B}_k := \sum_{m=1}^\nu [ \frac{\partial  B_k(\underline{\theta})}{\partial \theta_m}]^2,\\\nonumber
	\mathcal{C}_k &:= \sum_{m=1}^\nu [ \frac{\partial  C_k(\underline{\theta})}{\partial \theta_m}]^2,
	\quad \quad
	\mathcal{D}_{kl} := \sum_{m=1}^\nu [\frac{ \partial D_{kl}(\underline{\theta})}{\partial \theta_m}]^2,
\end{align}
through which we can express the precision $\epsilon$ of determining the full gradient vector as
\begin{align*}
	\epsilon^2	=  \mathcal{A} \var[ E^{(A)}]  
	&+ \sum_{k=1}^\nu \mathcal{B}_k \, \var[ E^{(B)}_k]
	+ \sum_{k=1}^\nu \mathcal{C}_k \, \var[ E^{(C)}_k]\\
	&+	\sum_{k=1}^\nu\sum_{l=k+1}^\nu \mathcal{D}_{kl} \,  \var[E^{(D)}_{kl}].
\end{align*}
We confirm the validity of the above error propagation formula in Fig.~\ref{check_variances_tapprox}.

Notice that the above equation is a sum over non-negative terms of the form
\begin{equation*}
	\epsilon^2 = \sum_{\mathbf{i} \in I} c_{\mathbf{i}} \var[x_{\mathbf{i}}],
\end{equation*}
where $I$ is an index set that indexes the terms in the above equation while 
$x_{\mathbf{i}}$ are statistical variables that correspond to coefficients in our
analytical approximation.  The coefficients $ c_{\mathbf{i}}$ are
given by, e.g., $\mathcal{B}_k$.
We can reduce $\epsilon^2$ by increasing the number of measurements that are used to determine, e.g., $E^{(C)}_k$.
In the following the variance $\var[x_{\mathbf{i}}]$ denotes the variance of a single measurement.
We distribute overall $N$ measurements optimally by assigning $N_\mathbf{i}$ measurements to estimating
the mean of the individual $x_{\mathbf{i}}$ variables as
\begin{equation*}
	N_\mathbf{i} = N \frac{\sqrt{c_\mathbf{i}  \var[x_{\mathbf{i}}]  } }{ T } = T \sqrt{c_\mathbf{i}  \var[x_{\mathbf{i}}]  } / \epsilon^2,
\end{equation*}
where we define
\begin{align}\label{tdefinition}
	T	:=& \sum_{\mathbf{i} \in I} \sqrt{c_\mathbf{i}  \var[x_{\mathbf{i}}]  } = \sqrt{\mathcal{A} \var[ E^{(A)}]} \\\nonumber
	 &+ \sum_{k=1}^\nu \sqrt{\mathcal{B}_k \, \var[ E^{(B)}_k]}
	+ \sum_{k=1}^\nu \sqrt{ \mathcal{C}_k \, \var[ E^{(C)}_k] }\\\nonumber
	&+	\sum_{k=1}^\nu\sum_{l=k+1}^\nu \sqrt{  \mathcal{D}_{kl} \,  \var[E^{(D)}_{kl}]  } .
\end{align}
Indeed the overall number of measurements is determined as $N = T^2/\epsilon^2$.
Furthermore, in this optimally distributed scheme the reduced individual variances are given by
\begin{equation}
	\var[x_{\mathbf{i}}]  / N_\mathbf{i}
	=
	\epsilon^2 \frac{   \var[x_{\mathbf{i}}]   }{   T \sqrt{c_\mathbf{i}  \var[x_{\mathbf{i}}]  } }
	=
	\epsilon^2  \frac{  \sqrt{\var[x_{\mathbf{i}}]  }  }{T  \sqrt{ c_\mathbf{i} }  }.
\end{equation}

Let us note that the quantity $T$ which determines our measurement cost depends on the parameters $\underline{\theta}$
and for example at $\underline{\theta} = \underline{0}$ we exactly obtain the measurement cost of the gradient vector as
\begin{align} \label{gradient_cost}
	T \rvert_{ \underline{\theta} = \underline{0} } =& \sum_{m=1}^\nu 
	\lvert \frac{\partial  B_m(\underline{\theta})}{\partial \theta_m} \rvert_{ \underline{\theta} = \underline{0} } \, \, \sigma[ E^{(B)}_m]\\
	=&  \sum_{m=1}^\nu  \sqrt{ \var[ E^{(B)}_m] } /2	 =: T_{grad}.
\end{align} 
Here we have used that $\frac{\partial  B_m(\underline{\theta})}{\partial \theta_m}\rvert_{ \underline{\theta} = \underline{0} } = 1/2$
and via $N_{grad} = T_{grad}^2 / \epsilon^2$, we exactly obtain the measurement cost of determining the gradient vector using parameter 
shift rules.

%-------------------------
\begin{figure*}[tb]
	\begin{centering}
		\includegraphics[width=0.5\textwidth]{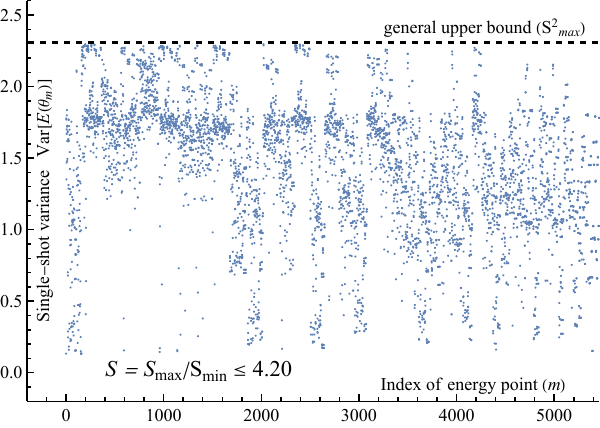}
		\includegraphics[width=0.4\textwidth]{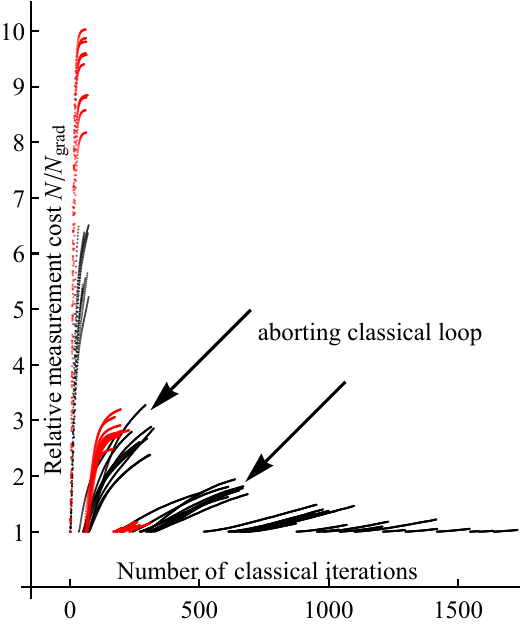}
		\caption{
			(Left)
			One needs to estimate the energy at shifted parameters $\underline{\theta}_m$
			in order to determine the coefficients in Eq.~\eqref{surfaceapprox}.
			We determine the corresponding single-shot variances $\var[E(\underline{\theta}_m)]$ in case of a 4-qubit spin-ring Hamiltonian
			in Eq.~\eqref{hamil-appendix} assuming that expectation values of Pauli strings
			are determined individually by sampling from the quantum computer.
			The single-shot estimation variance is generally upper bounded via Eq.~\eqref{smax},
			but it can be reduced significantly by applying more advanced techniques for
			simultaneously measuring commuting Pauli strings.
			Our \emph{relative} measurement cost depends on the ratio of minimal and maximal
			variances $S$ via Theorem~\ref{relative_cost}. In the present example we can estimate $S\leq4.2$ using $S^2_{min} \geq \min_{\theta} \var[E(\underline{\theta})]$.
			(Right)
			For all simulations of analytic descent from Fig.~2 in the main text we plot the \emph{exact} relative
			measurement cost $N/N_{grad}$ (which is upper bounded via Theorem~\ref{relative_cost})
			as a function of the classical iterations.
			In the initial evolutions $\delta$ is relatively large and analytic descent is therefore
			expensive. However, as we approach the optimum $\delta$ is smaller and the measurement
			overhead decreases and is guaranteed to vanish asymptotically. Sudden jumps in the plot indicate positions where we abort the
			classical internal loop and re-determine our classical approximation at the new reference point -- which costs exactly the same
			as determining the gradient vector. Red corresponds to the recompilation problem while black corresponds to the spin-ring Hamiltonian.
			\label{single-shot}
		}
	\end{centering}
\end{figure*}
%-------------------------

Let us summarise these results in the following theorem.
\begin{theorem}
	\label{theo:opt_dist}
	Let us denote variances of the single-measurement energy estimators as, e.g., $\var[ E^{(C)}_k]$. In order to determine the
	full gradient vector to a precision $\epsilon^2:=\sum_{k=1}^\nu \var[ \partial_m E(\underline{\theta}) ]$, 
	we need to distribute overall $N = T^2 / \epsilon^2$ measurements, where $T$ is defined in Eq.~\eqref{tdefinition}. When optimally distributed,
	estimating the coefficients in our analytical approximation requires the following
	number of measurements as
	\begin{align*}
		N^{(A)} &= T \sqrt{ \mathcal{A} \, \var[ E^{(A)}] }/\epsilon^2\\
		N^{(B)}_k &= T \sqrt{ \mathcal{B}_k \, \var[ E^{(B)}_k] }/\epsilon^2,\\
		N^{(C)}_k &= T \sqrt{ \mathcal{C}_k \, \var[ E^{(C)}_k] }/\epsilon^2,\\
		N^{(D)}_{kl} &= T \sqrt{ \mathcal{D}_{kl} \, \var[ E^{(D)}_{kl}] }/\epsilon^2,
	\end{align*}
	where, e.g., $N^{(C)}_k$ measurements are used to estimate the coefficient $E^{(C)}_k$.
	Here $\mathcal{A}$, $\mathcal{B}_k$, $\mathcal{C}_k$ and $ \mathcal{D}_{kl}$ are trigonometric
	polynomials defined in Eq.~\eqref{calCoeffs}.
\end{theorem}
It is important to recognise that the number of measurements, e.g., $N^{(A)}$, depend on the parameters $\underline{\theta}$,
however, this dependence is completely absorbed by the trigonometric polynomials, e.g., $\mathcal{A}(\underline{\theta})$. It follows that
we do not even need to explicitly/exactly know the coefficients, e.g., $E^{(A)}$ (only their variances which is significantly cheaper
to estimate) in order to determine how many measurements are required
to estimate the gradient vector via the analytic descent approach to a given precision. We provide an efficient
C code that exactly computes these trigonometric polynomials \cite{gitcode}.

\subsection{Measurement cost as a function of distance from the reference point}
While our classical algorithm evaluates the exact coefficients from Eq.~\eqref{calCoeffs}
via the efficient C code \cite{gitcode}, here we obtain local approximations of these
coefficients in order to be able to generally compare measurement costs of the
analytic descent approach to existing techniques.

Let us first expand $\mathcal{A}$ for small arguments $\theta_k$ as
\begin{equation} 	\label{cal_eqs}
	\mathcal{A} = \sum_{m=1}^\nu [\frac{\partial A(\underline{\theta}) }{\partial \theta_m}]^2 
		= ( 1+ \mathcal{O}(\nu \delta^2))^2	\sum_{m=1}^\nu \sin^2[\theta_m]/4   
\end{equation}
Above we have collected the leading terms in $\delta:=\lVert \underline{\theta}\rVert_\infty $ after expanding
the trigonometric functions for small arguments $\theta_k$ for
 $k \in \{1, 2, \dots, \nu\}$, such as, for example $\cos[\theta_k ] = 1 + \mathcal{O}(\delta^2)$.

We also expand the terms $\mathcal{B}_k$ and $\mathcal{C}_k$ as
\begin{align*}
	\mathcal{B}_k = \sum_{m=1}^\nu [ \frac{\partial  B_k(\underline{\theta})}{\partial \theta_m}]^2 
	&=  [ \frac{\partial  B_k(\underline{\theta})}{\partial \theta_k}]^2  + \mathcal{O}( \delta^4)\\
	&= 1/4 \times (1 + \mathcal{O}( \nu \delta^2))^2  \\
	\mathcal{C}_k = \sum_{m=1}^\nu [ \frac{\partial  C_k(\underline{\theta})}{\partial \theta_m}]^2
	&= [ \frac{\partial  C_k(\underline{\theta})}{\partial \theta_k}]^2  + \mathcal{O}( \delta^6)\\
	&= \sin^2[\theta_k]/4 \times (1+\mathcal{O}(\nu \delta^2 ) )^2.
\end{align*}
Finally, we expand the terms $\mathcal{D}_{kl}$ as
\begin{align*}
	\mathcal{D}_{kl} &= \sum_{m=1}^\nu [\frac{ \partial D_{kl}(\underline{\theta})}{\partial \theta_m}]^2
	= [\frac{ \partial D_{kl}(\underline{\theta})}{\partial \theta_k}]^2  {+}  [\frac{ \partial D_{kl}(\underline{\theta})}{\partial \theta_l}]^2 {+} \mathcal{O}( \delta^6)\\
	&= \Bigg( \sin^2[\theta_k]/16 + \sin^2[\theta_l]/16\Bigg) (1 + \mathcal{O}(\nu \delta^2))^2.
\end{align*}

Similarly, we can compute square roots of the above terms using the series expansion for $b$ smaller than
$a$ as $\sqrt{a + b} = \sqrt{a} + b/(2\sqrt{a}) + \dots$
as
\begin{align*}
	\sqrt{\mathcal{A}} &= \tfrac{1}{2}  \Vert \underline{\theta} \rVert_2 + \mathcal{O}(\nu \delta^3 ),
	\quad \quad
	\sqrt{\mathcal{B}_k} = \tfrac{1}{2}  + \mathcal{O}(\nu \delta^2 ),\\
	\sqrt{\mathcal{C}_k} &= \tfrac{1}{2}   \lvert \theta_k \rvert + \mathcal{O}(\nu \delta^3 ),
\,
	\sqrt{\mathcal{D}_{kl}} = \sqrt{( \theta_k^2 {+ }  \theta_l ^2  )}/4 + \mathcal{O}(\nu \delta^3 ).
\end{align*}
Let us now substitute these approximations back to the expression for $T$ as
\begin{align} \nonumber
	T	&=  \tfrac{1}{2}  \lVert \underline{\theta} \rVert_2 \,  \sigma[ E^{(A)}]
	+  \tfrac{1}{2} \sum_{k=1}^\nu  \sigma[ E^{(B)}_k] 
	+  \tfrac{1}{2}   \sum_{k=1}^\nu \lvert \theta_k \rvert   \, \sigma[ E^{(C)}_k] \\\nonumber
	&+	 \tfrac{1}{4} \sum_{k=1}^\nu\sum_{l=k+1}^\nu  \sqrt{ \theta_k ^2 +  \theta_l ^2 } \,  \sigma[E^{(D)}_{kl}] \\
	&+ \mathcal{O}(\nu \delta^2 )
	+ \mathcal{O}(\nu^2 \delta^3 ) \label{tapproximation}
\end{align}
and we have introduced the abbreviation $ \sigma[ \cdot ]  := \sqrt{\var[\cdot]}$. We verify the validity of this analytical approximation
in Fig.~\ref{check_variances_tapprox} and our numerical simulations confirm that the dominant error term scales
as  $\mathcal{O}(\delta^2 )$.

\subsection{Upper bounding single-shot variances and relative costs\label{theo2sec}}

Let us now state our result on bounding the measurement cost of analytic descent 
relative to the cost of a gradient evaluation.

\begin{theorem}\label{relative_cost}
	Let us introduce the ratio of the maximal single-shot variance for determining
	a single point on the energy surface $S^2 := \frac{  \max_{\underline{\theta}} \var[E(\underline{\theta})]  }  {  S^2_{min}  }$, relative to the minimal single-shot variance of determining a full gradient vector
	where we define $S^2_{min}$ in Eq.~\eqref{eq_smin}.
	The measurement cost of the analytic descent approach
	relative to determining a single gradient vector to the same precision $\epsilon$ is generally
	upper bounded as
	\begin{equation}
		N/N_{grad}  \leq [ 1 + \delta S (\sqrt{2} + \nu)]^2 	+ \mathcal{O}(\delta^2 ) + \mathcal{O}(\nu \delta^3 ).
	\end{equation}
	As such, the relative measurement cost scales as $1 + \mathcal{O}(\delta\nu)$ for small displacements.
	This also guarantees that asymptotically when approaching the optimum, and therefore $\delta \rightarrow 0$,
	the cost of analytic descent is the same as the cost of determining a gradient vector. 
\end{theorem}
\begin{proof}
Let us first introduce an upper bound on the single-shot variance $\var[E(\underline{\theta})] \leq S_{max}^2$
when estimating the energy via 
\begin{equation}\label{smax}
	S_{max}^2 := \max_{\underline{\theta}} \var[E(\underline{\theta})]
	\leq
	\sum_{k=1}^{r_H} |c_k|^2.
\end{equation}
This inequality provides a convenient, explicit formula in the specific case
when we can express the expected value
$E(\underline{\theta}) = \tr[\mathcal{H} \rho(\underline{\theta})]$
via the Hamiltonian
$\mathcal{H} = \sum_{k=1}^{r_H} |c_k|^2 P_k$
which decomposes into Pauli strings $P_k$. The above upper bound establishes that given the Pauli
decomposition of the Hamiltonian (as typical in practice) we can generally upper
bound the single-shot variances via \cite{van2021measurement} in terms of the coefficients.
Note that $S_{max}^2$ typically grows polynomially (via the number of Pauli terms $r_H$) with the number of qubits 
and $S_{max}^2$ can be significantly reduced by optimally distributing measurements or
by simultaneously measuring commuting Pauli terms via advanced techniques
\cite{Crawford2019}. We illustrate the upper bound and the actual single-shot energy
variances in Fig.~\ref{single-shot}.

Using the single-shot variance upper bound above, it follows that the standard deviations of estimating
our coefficients are upper bounded as $\sigma[ E^{(A)}] \leq S_{max}$,
$\sigma[ E^{(C)}_k] \leq S_{max}$,
$\sigma[E^{(B)}_k] \leq \sqrt{2} S_{max}$
and $ \sigma[E^{(D)}_{kl}] \leq 2 S_{max}$. 
It may be useful in practice that we can explicitly upper bound the measurement cost $T$ 
from Eq.~\eqref{tapproximation} as
\begin{align*}
	T	\leq&  \tfrac{1}{2}  \lVert \underline{\theta} \rVert_2 S_{max}
	+   \tfrac{1}{\sqrt{2}} \nu S_{max}
	+  \tfrac{1}{2}   \lVert \underline{\theta} \rVert_1 S_{max}\\
	&+	  \sum_{k=1}^\nu\sum_{l=k+1}^\nu  \sqrt{  \theta_k ^2 +   \theta_l ^2 } S_{max}
	+ \mathcal{O}(\nu \delta^2 )
	+ \mathcal{O}(\nu^2 \delta^3 ).
\end{align*}
Instead of upper bounding the cost of analytic descent as above, let us now derive an explicit upper bound on the measurement cost of
the analytic descent approach \emph{relative to the cost of determining a gradient vector.}

\begin{widetext}

For this reason we first compute the ratio
\begin{equation}
	\frac{T}{T_{grad}  }	= 1 + \frac{
		\lVert \underline{\theta} \rVert_2 \,  \sigma[ E^{(A)}]
		+    \sum_{k=1}^\nu \lvert \theta_k \rvert   \, \sigma[ E^{(C)}_k] 
		+	 \tfrac{1}{2} \sum_{k=1}^\nu\sum_{l=k+1}^\nu  \sqrt{ \theta_k ^2 +  \theta_l ^2 } \,  \sigma[E^{(D)}_{kl}] 
		+ \mathcal{O}(\nu \delta^2 )
		+ \mathcal{O}(\nu^2 \delta^3 )
	}{ \sum_{m=1}^\nu  \sigma[ E^{(B)}_m] },
\end{equation}
where we have used that $T_{grad} = \sum_{m=1}^\nu  \sigma[ E^{(B)}_m]  /2$ and in the nominator we have 
	used our asymptotic approximation of $T$ from Eq.~\eqref{tapproximation}.
The denominator is generally lower bounded as $\sum_{m=1}^\nu  \sigma[ E^{(B)}_m]  \geq \sqrt{2} \nu S_{min}$
and thus allows us to obtain the upper bound as
\begin{equation}\label{upper_bound_eq}
	\frac{T}{T_{grad}  }	\leq 1 + \frac{
		\lVert \underline{\theta} \rVert_2 S_{max}
		+   \lVert \underline{\theta} \rVert_1   S_{max}
		+	 \sum_{k=1}^\nu\sum_{l=k+1}^\nu  \sqrt{ \theta_k ^2 +  \theta_l ^2 } S_{max}
		+ \mathcal{O}(\nu \delta^2 )
		+ \mathcal{O}(\nu^2 \delta^3 )
	}{ \sqrt{2} \nu S_{min} }.
\end{equation}
\end{widetext}
	Above we have defined a minimal single-shot variance as averaged over parameter shifts
	\begin{align}\label{eq_smin}
		S_{min} &:= \min_{\underline{\theta}} \tfrac{1}{\nu \sqrt{2}}\sum_{m=1}^\nu \sqrt{Q},\\
		\text{with} \quad \quad
		Q &:= \var[E(\underline{\theta} + \tfrac{1}{2}\pi \underline{v}_m )]
		+
		\var[E(\underline{\theta} - \tfrac{1}{2}\pi \underline{v}_m )].  \nonumber
	\end{align}
	Here $\underline{v}_m$ are the standard basis vectors in parameter space.
	This minimal single-shot variance is generally lower bounded as
	$S^2_{min} \geq \min_{\theta} \var[E(\underline{\theta})]$, however,
	note that the variance of the energy measurement can vanish as $\var[E(\underline{\theta})] \rightarrow 0$, e.g.,
	when approaching an eigenstate of a diagonal problem Hamiltonian.
	In case of such systems we need to use our general definition of
	$S_{min}$ which indeed cannot vanish as $S_{min}>0$ except for trivial
	problem definitions that are not relevant in practice, e.g., all quantum gates in the
	ansatz, the problem Hamiltonian and the quantum state $\rho$ are diagonal in
	the same basis.

In the nominator of Eq.~\eqref{upper_bound_eq} the individual terms are upper bounded as
\begin{align*}
	&\lVert \underline{\theta} \rVert_2 /(\sqrt{2} \nu) \leq \delta/\sqrt{2},
	\quad \quad
	\lVert \underline{\theta} \rVert_1/(\sqrt{2} \nu) \leq \delta/\sqrt{2},\\
	&\sum_{k=1}^\nu\sum_{l=k+1}^\nu  \sqrt{ 	 \theta_k ^2 +   \theta_l^2 } /(\sqrt{2} \nu) \leq  \delta  \nu.
\end{align*} 
It follows that the measurement cost of the analytic descent approach relative to determining the gradient
vector is
\begin{align}
	N/N_{grad}  &	= [T/T_{grad}]^2  \\\nonumber
	&\leq  [ 1 + \delta S (\sqrt{2} + \nu)]^2 	+ \mathcal{O}(\delta^2 ) + \mathcal{O}(\nu \delta^3 ),
\end{align}
where we have introduced the abbreviation for the ratio of lower and upper bounds $S := S_{max}/S_{min}$.
\end{proof}

\subsection{Symmetry of the energy surface around the optimum\label{symmetry}}
At a local optimum one finds that the gradient vanishes as $g_m = 0$ for $m \in \{1, \dots, \nu\}$.
We set $ \underline{\theta}_0 := \underline{\theta}_{opt}$ and therefore $\underline{\theta}=0$.
The explicit form of the leading terms in the energy surface can be expressed as
\begin{align}\nonumber
	E(\underline{\theta}) :=& \tr[\mathcal{H} \,  \Phi(\underline{\theta})\rho_0]
	=A(\underline{\theta})  E^{(A)}   + \sum_{k=1}^\nu [C_k(\underline{\theta}) E^{(C)}_k] \\
	&+
	\sum_{k}^\nu\sum_{l=k+1}^\nu [D_{kl}(\underline{\theta}) E^{(D)}_{kl}] + \mathcal{O}(\sin^3 \delta).
\end{align}
and we have used that $E^{(B)}_k  = 0$ due to $g_k = 0$. We now make two observations which
pose strict constraints on the geometry of the energy surface around local optima.
First, the energy function in this case is (approximately) reflection symmetric via 
\begin{equation}
	E(\underline{\theta}) = E(-\underline{\theta}) +  \mathcal{O}(\sin^3 \delta),
\end{equation}
due to the reflection symmetry of the basis functions $A(\underline{\theta})$, $C_k(\underline{\theta})$
and $D_{kl}(\underline{\theta})$.
Second, any slice of the energy function is just a shifted cosine function
as 
\begin{equation*}
	E(\theta_k \underline{v}_k) = E^{(A)} (1+\cos[\theta_k])/2  + E^{(C)}_k (1-\cos[\theta_k])/2,
\end{equation*}
which can be written as $a + b \cos(\theta_k) $  and $a =  E^{(A)} + E^{(C)}_k $, 
while $b =  E^{(A)} - E^{(C)}_k $.

\subsection{Relation to the Hessian matrix and to a Taylor expansion \label{hessianSection}}

One can show that the coefficients used to determine our
approximation of the energy surface are related to partial derivatives of the energy surface. In particular, the gradient vector $g_m$ from
Eq.~\eqref{full-grad-equation} can be expressed exactly at the point $\underline{\theta}$ as
\begin{align} \label{param-shift-rule}
	g_m &= [\partial_m E(\underline{\theta})] |_{\underline{\theta} = \underline{0}} = 
	E^{(B)}_m [\frac{\partial  B_m(\underline{\theta})}{\partial \theta_m}]|_{\underline{\theta} = \underline{0}}\\
	&= E^{(B)}_m/2
	= [E(\tfrac{1}{2}\pi \underline{v}_k) -   E(-\tfrac{1}{2}\pi \underline{v}_k)]/2.\nonumber
\end{align}
This is the well-known parameter-shift rule, which estimates the gradient via sampling the energy function
at two different points \cite{paramshift}.

\begin{widetext}
The mixed second partial derivatives can similarly be expressed exactly using Eq.~\eqref{surfaceapprox} as
\begin{align}
	[\partial_m \partial_n E(\underline{\theta})] |_{\underline{\theta} = \underline{0}} 
	&=
	E^{(D)}_{kl} [\frac{ \partial^2 D_{kl}(\underline{\theta})}{\partial \theta_m \partial \theta_n} ]|_{\underline{\theta} = \underline{0}} 
	= E^{(D)}_{kl}/4\\\nonumber
	&=
	[E(\tfrac{1}{2}\pi \underline{v}_k + \tfrac{1}{2}\pi \underline{v}_l )
	+  E(-\tfrac{1}{2}\pi \underline{v}_k - \tfrac{1}{2}\pi \underline{v}_l )
	-	  E(-\tfrac{1}{2}\pi \underline{v}_k + \tfrac{1}{2}\pi \underline{v}_l )
	-  E(\tfrac{1}{2}\pi \underline{v}_k - \tfrac{1}{2}\pi \underline{v}_l )]/4,
\end{align}
when $n \neq m$ and
\begin{align}
	[\partial_m \partial_m E(\underline{\theta})] |_{\underline{\theta} = \underline{0}} 
	&=
	E^{(A)} [\frac{ \partial^2 A(\underline{\theta})}{\partial \theta_m \partial \theta_m} ]|_{\underline{\theta} = \underline{0}} 
	+E^{(C)}_{m} [\frac{ \partial^2 C_{m}(\underline{\theta})}{\partial \theta_m \partial \theta_m} ]|_{\underline{\theta} = \underline{0}} \\\nonumber
	&=[E^{(C)}_{m} - E^{(A)}]/2 =E(\pi \underline{v}_k) - E(\underline{0}).
\end{align}
\end{widetext}

%-------------------------
\begin{figure*}[tb]
	\begin{centering}
		\includegraphics[width=0.9\textwidth]{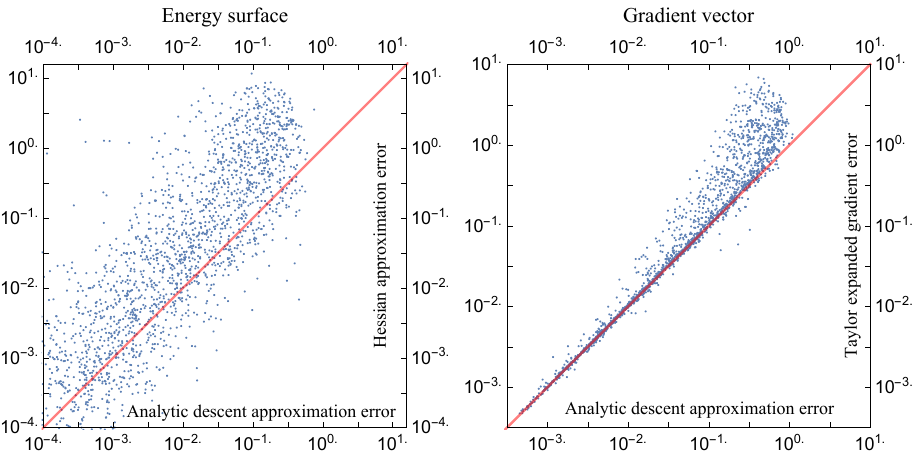}
		\caption{
			Approximating the energy surface $E(\underline{\theta})$ and the gradient vector $\underline{g}(\underline{\theta})$
			at randomly generated points $\underline{\theta}$ around the ground state of the spin-ring Hamiltonian from Sec.~\ref{simulations}
			using analytic descent and using the Taylor expansion from
			Eq.~\eqref{taylor_expansion}. Approximation error of the gradient is computed via the vector distance $\lVert \underline{v}-\underline{g} \rVert_{\infty}$.
			Red line in the diagonal corresponds to the case when the two approaches give the same error.
			Although both the Taylor expansion and analytic descent have the same asymptotic scaling in $\delta$,
			analytic descent typically significantly outperforms the Taylor expansion (sometimes by as much as 2 orders
			of magnitude) for non-vanishing $\delta$---as
			relevant in practice. 
			\label{hessian_plot}
		}
	\end{centering}
\end{figure*}
%-------------------------	

To conclude, we express explicitly elements of the gradient vector as
\begin{equation}
	g_m =  [\partial_m E(\underline{\theta})] |_{\underline{\theta} = \underline{0}} = E^{(B)}_m/2
\end{equation}
and elements of the Hessian matrix as
\begin{equation}\label{hessian_matr}
	H_{mn} =  [\partial_m \partial_n E(\underline{\theta})] |_{\underline{\theta} = \underline{0}}  =
	\begin{cases} 
		[E^{(C)}_{m} - E^{(A)}]/2
		& \mbox{if } 	m = n\\
		E^{(D)}_{mn}/4
		& \mbox{if } 	m \neq n.\\
	\end{cases} 
\end{equation}	
This means that when querying the energy function in Sec.~\ref{energySection} and in Sec.~\ref{gradientSection},
the information we determine is very closely related to the Hessian and the gradient of the energy surface.
As such, when not considering shot noise, we require the same quantum resources to determine both the analytic descent approximation
and the well-known Taylor expansion as
\begin{equation} \label{taylor_expansion}
	E(\underline{\theta}) = E(\underline{0}) + \underline{\theta} \, \underline{g} + \tfrac{1}{2}  \underline{\theta}^T \, H_{mn} \, \underline{\theta}
	+ \mathcal{O}(\delta^3),
\end{equation}
which has the same asymptotic scaling in $\delta$ as the analytic descent approach.
Let us now explain why the analytic descent approach may be preferable.
First, the Taylor expansion is a quadratic polynomial in the variables $\theta_k$, i.e., it contains
no degree-3 contribution. In contrast, the analytic descent approach is an infinite-degree polynomial
in the variables $\theta_k$, i.e., an analytic function, as it is composed of trigonometric functions, such
as $\cos(\theta_k)$. Even though in the limit when $\theta_k \rightarrow 0$ for all $k$  asymptotically
both approaches
have the same approximation errors, in practice one always aims to use the approximations for finite,
non-vanishing parameters $\theta_k$. We compare approximation errors in case of analytic descent and 
in case of the Taylor expansion in Fig.~\ref{hessian_plot} (left). Indeed, our trigonometric expansion typically gives a better approximation of the energy surface (majority of points above the red line) and in some cases the approximation errors are about 2 orders of magnitude smaller.

Let us consider a specific example that nicely illustrates how our trigonometric  
approximation outperforms the above Taylor expansion. Let us assume that we move away from the reference
point only along a single variable $\theta_k$. In this case our trigonometric series is \emph{exact} for arbitrarily large $\theta_k$, however,
the Taylor expansion breaks down and its error increases infinitely: in the extreme scenario when $\theta_k=10^{10}$ the approximation error can be as large
as $10^{10}$ while our 	trigonometric series is exact.
The argument approximately holds even when we move along every parameter but there are a few
dominant components, e.g., $\theta_k$, $\theta_{k+1}$ etc. This can often happen in practice.
This illustrates that while the Taylor expansion only captures the energy surface locally,
analytic descent also captures some of the global features too.

Of course, in our optimisation algorithm we do not actually use the approximation of the energy energy surface,
but	instead the resulting analytical gradient vector. It is therefore more meaningful to compare how the gradient
vector can be approximated by the two techniques. Our optimisation technique is compatible with a (linear) Taylor expansion
as the analytical gradients could be used in our algorithm
\begin{equation*}
	\partial_m 	E(\underline{\theta}) = \underline{g}(\underline{0}) + \tfrac{1}{2} \sum_{n=1}^\nu  H_{mn} \theta_m + \mathcal{O}(\delta^2).
\end{equation*}	
When not considering shot noise, the resulting approach would require the same quantum resources as
the trigonometric series, i.e., same number of coefficients determined,  but would require reduced classical
computational resources.
Nevertheless, we assume that the classical computational resources required for computing the trigonometric
series are free, and therefore we prefer to use the trigonometric series.
In Fig.~\ref{hessian_plot} we compare these gradient
approximation 	errors and find that the superiority of the trigonometric series is even more pronounced:
analytic descent almost always outperforms the Taylor expansion as almost all dots are above the red line.

\section{Expanding the metric tensor entries \label{metric-tensor-sec}}

It was shown in \cite{koczor2019quantum} that the quantum Fisher information matrix can be
approximated by the scalar product
\begin{equation}
	[\qf]_{mn} = 2 \tr[\frac{\partial \rho(\underline{\theta})}{\partial_m} \frac{\partial \rho(\underline{\theta})}{\partial_n}],
\end{equation}
which relation becomes exact in the limit of pure states.
Here we have denoted $\rho(\underline{\theta}):= \Phi(\underline{\theta})\rho_0$. We can straightforwardly express
the partial derivatives via the partial derivative of the mapping
\begin{equation}
	\frac{\partial \rho(\underline{\theta})}{\partial_m}  = \frac{\partial \Phi(\underline{\theta})}{\partial_m} \rho_0
	= \frac{\partial \tilde{\Phi}(\underline{\theta})}{\partial_m} \rho_0 +  \mathcal{O}(\sin^3 \delta),
\end{equation}
which we aim to express explicitly using our approximate mapping $\tilde{\Phi}(\underline{\theta})$
from Eq.~\eqref{approx-mapping}. We can compute the derivative analytically as
\begin{align}\nonumber
	\frac{\partial \tilde{\Phi}(\underline{\theta})}{\partial_m}
	=& \frac{\partial A(\underline{\theta}) }{\partial \theta_m} \Phi^{(A)}   + \sum_{k=1}^\nu [ \frac{\partial  B_k(\underline{\theta})}{\partial \theta_m} \Phi^{(B)}_k + \frac{ \partial C_k(\underline{\theta}) }{\partial \theta_m} \Phi^{(C)}_k]\\
	 &+
	\sum_{k}^\nu\sum_{l=k+1}^\nu [\frac{ \partial D_{kl}(\underline{\theta})}{\partial \theta_m} \Phi^{(D)}_{kl}] + \mathcal{O}(\sin^2\delta).
	\label{mapping-deriv-approx}
\end{align}
and note that this expression is directly analogous to the gradient vector from Eq.~\eqref{full-grad-equation}, and
we have defined the partial derivatives of the monomials, such as $\frac{\partial A(\underline{\theta}) }{\partial \theta_m}$,
in Sec.~\ref{gradientSection}. Expanding the quantum Fisher information to leading terms only
results in
\begin{equation*}
	[\qf]_{mn} =
	\mathcal{F}_{BB} F_{BB}(\underline{\theta})
	+ \mathcal{F}_{AB} F_{AB}(\underline{\theta}) + \dots + \mathcal{O}(\sin^2\delta).
\end{equation*}
We do not write out all the terms explicitly for clarity -- however, note that they
could be computed straightforwardly.

Similarly as before, we have monomials that completely absorb the
continuous dependence on the parameters $\underline{\theta}$ and their explicit forms can be computed as
\begin{align}
	F_{BB}(\underline{\theta}) :=& 2\frac{\partial  B_m(\underline{\theta})}{\partial \theta_m}  \frac{\partial  B_n(\underline{\theta})}{\partial \theta_n}\\
	F_{AB}(\underline{\theta}) :=&  2\frac{\partial B_m(\underline{\theta}) }{\partial \theta_m}  \frac{\partial A(\underline{\theta}) }{\partial \theta_n} 
	+ 2 \frac{\partial A(\underline{\theta}) }{\partial \theta_m}  \frac{\partial B_n(\underline{\theta}) }{\partial \theta_n}
\end{align}
These functions multiply the
coefficients, e.g., $\tr[(\Phi^{(B)} \rho_0) (\Phi^{(B)} \rho_0)]$, which
can be computed via the discrete transformations as
\begin{align*}
	\mathcal{F}_{BB} =& \tr[(\Phi^{(B)} \rho_0) (\Phi^{(B)} \rho_0)]\\
	 =&+	\tr[\rho(\tfrac{1}{2}\pi \underline{v}_k) \rho(\tfrac{1}{2}\pi \underline{v}_k)]
	+\tr[\rho(-\tfrac{1}{2}\pi \underline{v}_k) \rho(-\tfrac{1}{2}\pi \underline{v}_k)]\\
	&-\tr[\rho(-\tfrac{1}{2}\pi \underline{v}_k) \rho(\tfrac{1}{2}\pi \underline{v}_k)]
	-\tr[\rho(\tfrac{1}{2}\pi \underline{v}_k) \rho(-\tfrac{1}{2}\pi \underline{v}_k)]\\
	\mathcal{F}_{AB} =&  \tr[(\Phi^{(A)} \rho_0) (\Phi^{(B)} \rho_0)]\\
	 =& 
	\tr[\rho(\underline{0}) \rho(\tfrac{1}{2}\pi \underline{v}_k)] - \tr[\rho(\underline{0}) \rho(-\tfrac{1}{2}\pi \underline{v}_k)].
\end{align*}
The coefficients therefore can be estimated by estimating the
overlap between the states, as e.g., $\rho(\underline{0})$ and $\rho(\tfrac{1}{2}\pi \underline{v}_k)$.
These can be straightforwardly estimated using SWAP tests or, in the case of pure states,
using Hadamard tests as, e.g., 
\begin{equation}
	\tr[\rho(\underline{0}) \rho(\tfrac{1}{2}\pi \underline{v}_k)] = |\langle \psi(\underline{0}) | \psi(\tfrac{1}{2}\pi \underline{v}_k)\rangle|^2.
\end{equation}

\section{Numerical computations}
\subsection{Classical algorithm for computing the gradient vector \label{classical-alg-sec}}

%-------------------------
\begin{figure}[tb]
	\begin{centering}
		\includegraphics[width=0.48\textwidth]{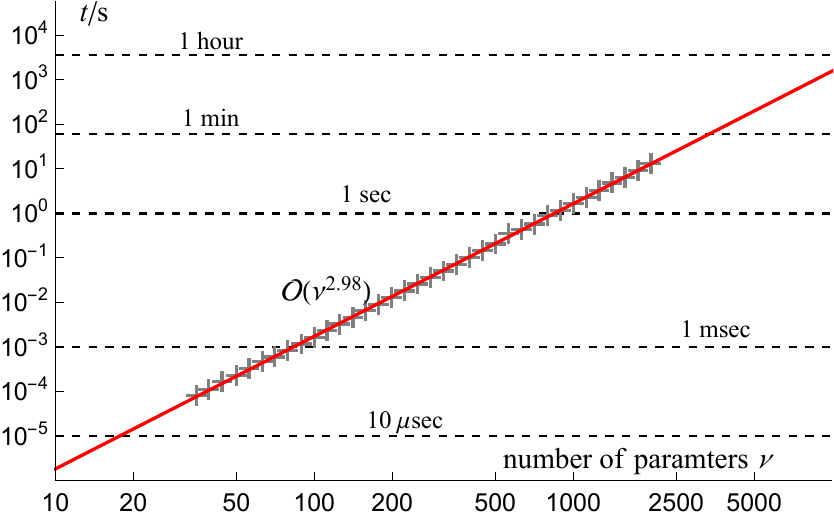}
		\caption{
			Classically computing the gradient vector using our efficient C
			code \cite{gitcode}. Execution times estimated on a laptop for
			an increasing number of parameters $\nu$ confirm the theoretical
			complexity $\mathcal{O}(\nu^3)$ from Sec.~\ref{classical-alg-sec}.
			Descending 1000 steps towards the minimum of the classical
			function can be performed in a matter of minutes for up to many
			hundreds of parameters. Our code was executed on a single thread, but the algorithm
			described in Sec.~\ref{classical-alg-sec} could be parallelised.
			\label{figtimes}
		}
	\end{centering}
\end{figure}
%-------------------------

We now describe how the analytical gradient from
Eq.~\eqref{full-grad-equation} can be computed classically
efficiently. We assume the coefficients 
$E^{(A)},E^{(B)}_k,E^{(C)}_k, E^{(D)}_{kl}$ are already
determined and accessible in RAM. This requires
$\mathcal{O}(\nu^2)$ space, which is reasonable
for up to thousands of parameters. 

First our classical algorithm needs to compute the monomials, e.g., 
$ \frac{\partial A(\underline{\theta}) }{\partial \theta_m}$
for a given input vector $\underline{\theta}$. We do this 
by precomputing and storing the basis functions 
$a(\theta_k), b(\theta_k) = 1\pm\cos(\theta_k)$ and $c(\theta_k) = \sin(\theta_k)/2$ and
\begin{align*}
	\frac{\partial a(\theta_k)}{\partial \theta_k} &= - \sin[\theta_k]/2, \quad \quad
	\frac{\partial b(\theta_k)}{\partial \theta_k}=  \cos[\theta_k]/2,\\
	\frac{\partial c(\theta_k)}{\partial \theta_k}&=  \sin[\theta_k]/2,
\end{align*}
for all parameters $k \in \{1, \dots \nu \}$.
This can be evaluated in $\mathcal{O}(\nu)$ time
and requires $\mathcal{O}(\nu)$ storage space.

In the next step we multiply together the basis functions $a(\theta_k)$ to obtain
$A(\underline{\theta})$ as
\begin{equation*}
	A(\underline{\theta}) = a(\theta_1) a(\theta_2) \cdots a(\theta_\nu),
\end{equation*}
and we store it.
All other monomials are obtained from this just
by dividing it by, e.g., $a(\theta_k)$, and
then multiplying it with, e.g., $\frac{\partial b(\theta_k)}{\partial \theta_k}$, which components we have already
precomputed. For example, the monomial $\frac{\partial B_k(\underline{\theta})}{\partial \theta_m}$
is obtained as
\begin{equation*}
	\frac{\partial B_k(\underline{\theta})}{\partial \theta_m} = \frac{ A(\underline{\theta}) }{ a(\theta_k) a(\theta_m) } \frac{\partial b(\theta_m)}{\partial \theta_m} b(\theta_k),
\end{equation*}
when $k \neq m$ and note that we have already
precomputed all components in the above equation.
In conclusion, evaluating all $\nu Q=\mathcal{O}(\nu^3)$ basis functions in the gradient vector
for a given input vector $\underline{\theta}$ can be done in $\mathcal{O}(\nu^3)$ time and
requires $\mathcal{O}(\nu^2)$ storage. We have estimated execution times of our C implementation
from \cite{gitcode} which confirms this theoretical complexity, refer to
Fig.~\ref{figtimes}. 

%-------------------------
\begin{figure*}[tb]
	\begin{centering}
		\includegraphics[width=0.6\textwidth]{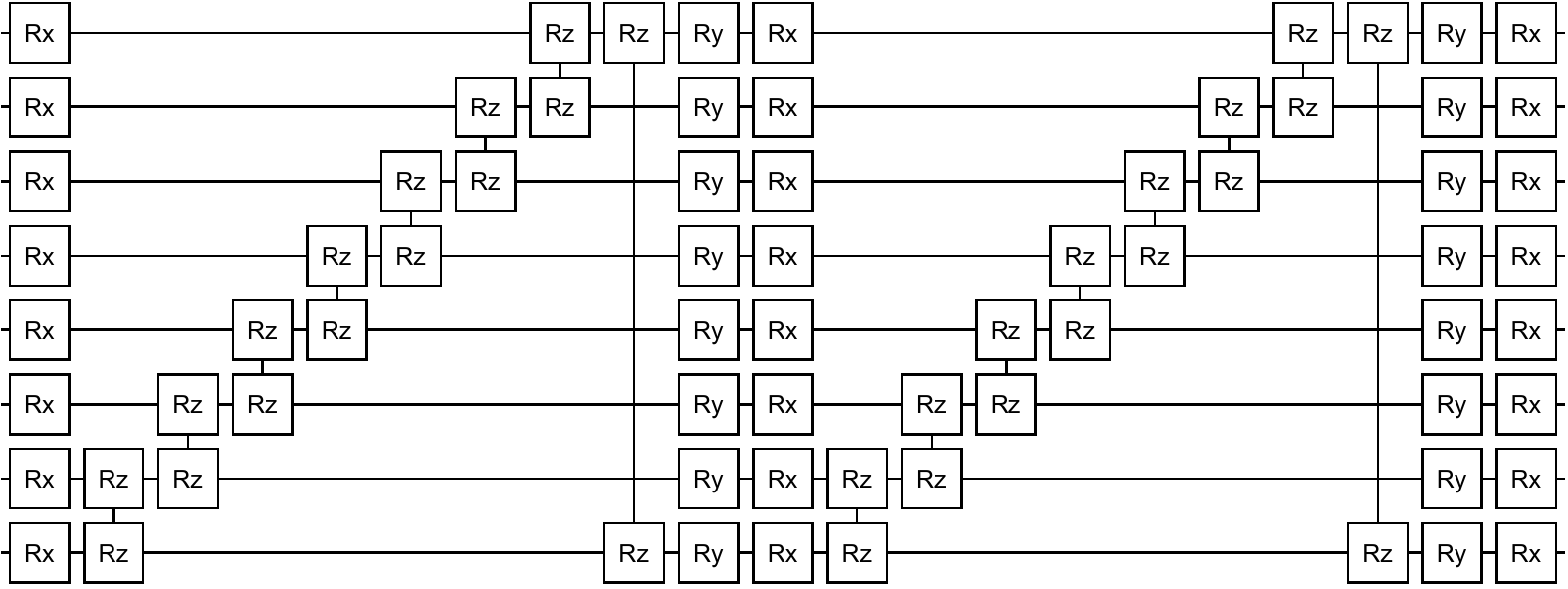}
		\caption{
			Example of a 2-block ansatz circuit of 8 qubits.
			We used 4-block circuits in our simulations.
			\label{figansatz}
		}
	\end{centering}
\end{figure*}
%-------------------------

\subsection{Classical algorithm for computing the optimal measurement distribution \label{classical-alg-measurement}}
Recall from Sec.~\ref{sec:opt_measurement} that the gradient variance can be expressed as
\begin{align}	\label{grad_variance}
	\epsilon^2	=&  \mathcal{A}(\underline{\theta}) \var[ E^{(A)}]  
	+ \sum_{k=1}^\nu \mathcal{B}_k(\underline{\theta}) \, \var[ E^{(B)}_k]\\\nonumber
	&+ \sum_{k=1}^\nu \mathcal{C}_k(\underline{\theta}) \, \var[ E^{(C)}_k]
	+	\sum_{l>k} \mathcal{D}_{kl} (\underline{\theta}) \,  \var[E^{(D)}_{kl}],
\end{align}
where $\mathcal{A}$, $\mathcal{B}_k$, $\mathcal{C}_k$ and  $\mathcal{D}_{kl}$
are trigonometric polynomials that depend on the parameters $\underline{\theta}$.
The variances, such as $\var[ E^{(A)}] $, are proportional to single-shot variances
of estimating the energy expectation values. We therefore assume that the experimentalist
has explicit knowledge of these (these can be estimated efficiently experimentally). The optimal measurement distribution can therefore be
obtained via Theorem~\ref{theo:opt_dist} by explicitly computing the above trigonometric
polynomials.

The analytical forms of these trigonometric polynomials are defined in Eq.\eqref{cal_eqs}:
they are sums of squares of the monomial derivatives as, e.g., $\frac{\partial B_k(\underline{\theta})}{\partial \theta_m}$, 
with respect to the index $m$.
As such, we only need to sum up the squares of these trigonometric monomials, which
our classical algorithm from Sec.~\ref{classical-alg-sec} computes in $\mathcal{O}(\nu^3)$ time.
In summary we require $\mathcal{O}(\nu^3)$ time and  $\mathcal{O}(\nu^2)$ storage for determining the
optimal measurement distribution. We have made available our efficient C code online \cite{gitcode}.

\subsection{Simulations in main text}
In Fig.~2 in the main text we considered two specific problems that aim to find the ground state of
a Hamiltonian. In the first case we consider a recompilation problem whereby we aim to recompile a 4-qubit
quantum circuit $\mathrm{C}_{4}[\mathrm{SWAP}_{12}] \mathrm{C}_{4}[\mathrm{SWAP}_{23}]$ that contains two
consecutive controlled-SWAP operators as relevant in the context of error mitigation \cite{endo2020hybrid, koczor2021exponential,koczor2021dominant, huggins2020virtual}. The recompilation
requires overall an $8$-qubit circuit which is initialised by entangling every qubit in the 4-qubit register
with qubits in an ancillary 4-qubit register as described in \cite{khatri2019quantumassisted}. Our $4$-qubit ansatz circuit consists
of $124$ parametrised quantum gates as illustrated in Fig.~\ref{figansatz} and we aim to optimise parameters of this
circuit such that the ground state of the Hamiltonian $- \sum_{k=1}^8 Z_k$ is found.

In the second scenario we consider the spin-ring Hamiltonian
\begin{equation} \label{hamil-appendix} 
	\sum_{i=1}^{N} J [ 
	X_{i}  X_{i+1}
	+ Y_{i} Y_{i+1} 
	+ Z_{i} Z_{i+1} 
	]
	+\sum_{i=1}^N  \omega_i \, Z_i,
\end{equation}
in which we have set $N+1=1$ and $X$, $Y$ and $Z$ are Pauli matrices. We randomly generate $-1 \leq \omega_i \leq 1$ and set $J=0.1$. 
Our $8$-qubit ansatz consists of $104$ parametrised quantum gates as illustrated in Fig.~\ref{figansatz}.

We simulate four different optimisers and estimate the level of quantum resources required to reach a certain
precision with respect to the exact ground-state energy. For this reason, we have determined the optimal
set of parameters $\underline{\theta}_{opt}$ and we initialise the optimisation in its vicinity:
we disturb the optimal parameters $\theta_k$ by adding uniformly randomly generated numbers in the
range $(-0.05,0.05)$. We estimate measurement costs by assuming that a single call to the quantum subroutine
determines the coefficients, such as $E_k^{(B)}$, to unit variance as $\var[E_k^{(B)}]=1$, and we count the overall
number of calls $N_s$ at every iteration. We simulate shot noise in all cases by adding Gaussian distributed
random numbers to the coefficients;
the standard deviation is related to the number of shots $N_E$ that we use to estimate a single coefficient
as $\sigma_E = 1/\sqrt{N_E}$.
Let us now detail how we set hyperparameters of each optimisation technique
such that they all can consistently reach a precision $\Delta E = 10^{-4}$ in determining the ground-state
energy.

\textbf{Simple gradient descent:}
Recall that a simple gradient descent update rule is defined as $\underline{\theta}_{k+1} = \underline{\theta}_{k} - \lambda \underline{g}$,
where $\underline{g}$ is the gradient vector
and in the following we refer to $\lambda$ as the step size.
We set the largest stable step size as $0.2$ and we set a fixed
precision of determining the full gradient vector as the Euclidean distance
$\epsilon^2:=\sum_{k=1}^\nu \var[ \partial_m E(\underline{\theta}) ]$. Recall that gradient descent
is guaranteed to converge in principle under an arbitrarily small precision \cite{sweke2019stochastic}, and therefore we
set a relatively low, constant precision $\epsilon^2 = 10^{-5}$ such that evolution approaches the convergence criterion 
with approximately a uniform convergence rate. We use the parameter shift rule from Eq.~\eqref{param-shift-rule} and thus
the measurement cost of a single iteration can be computed via  Eq.~\ref{gradient_cost} as
$N_s  = \epsilon^{-2} T_{grad}^2  = \epsilon^{-2} \nu^2/4$ given the single-shot variance $\var[E_k^{(B)}]=1$.
Note that while the measurement cost of a single \emph{iteration} is $N_s$,
	the number of shots to determine one of the $E_k^{(B)}$ coefficients is
$N_E = N_s/\nu = \epsilon^{-2} \nu/4$, i.e., for the spin-ring Hamiltonian we have used
$N_E = 10^{6.41}$ while for the recompilation problem we used $N_E = 10^{6.49}$ shots for
determining a single coefficient.
One can certainly increase the efficiency of simple gradient descent by adaptively setting the
gradient precision \cite{kubler2019adaptive} or using ADAM or SPSPA variants. However, we stress that
in order to be able to compare vastly different optimisation techniques and their convergence rates
we decided to set a constant precision. Of course, all other techniques would certainly benefit from
more advanced adaptive strategies, but this is beyond the scope of the present work.

\textbf{Analytic descent:} We set a small step size $0.01$ such that our classical gradient descent optimisation
follows a smooth evolution path, i.e., we assume that classical computation is free. Furthermore, we
use the same relatively low precision of determining the classical approximation to the gradient vector as in case of simple
gradient descent as $\epsilon^2 = 10^{-5}$. As such, the optimally distributed measurement cost of determining
our classical approximation at a reference point $\underline{\theta}_{0}$ is exactly the same as in case of simple
gradient descent as $N_s = \epsilon^{-2} \nu^2/4$. This measurement cost is slightly increased by a small
factor as we move away from the reference point since we need to collect further samples using the quantum
computer via our optimally distributed measurement scheme as illustrated in Fig.~\ref{single-shot}(right).
The measurement cost is a function of the parameters
as $N_s:= N_s(\underline{\theta})$ and we approximate the overall cost of a single iteration as the maximum of
this function. We find that in the early evolution, where analytic descent is less
beneficial, this measurement cost is at most by a factor of $10$ more expensive than the cost of determining 
a single gradient vector to the same precision -- since here the optimiser takes large jumps and the measurement cost
grows with the size of the jump. These measurement costs are explicitly shown in Fig.~\ref{single-shot}(b).
It is also evident from Fig.~\ref{single-shot}(b) that in the later evolutions the cost of an analytic-descent
iteration is only by a small factor $\leq 2$ more expensive than determining a gradient vector.
Note that in case of analytic descent the number of shots $N_E$ to determining single coefficients, as e.g.
	$N^{(C)}_k$,
	are distributed optimally via Theorem~\ref{theo:opt_dist} and thus cannot it be compared to that of other techniques'
	sampling rates.

In case of analytic descent we have aborted the internal classical optimisation loop if the exact energy, as determined via
a quantum computer, was increased. In a later section we explore an abort condition based on our similarity measure $f$.

\textbf{Hessian-based optimisation:} We determine the Hessian matrix from Eq.~\eqref{hessian_matr}
and the gradient vector Eq.~\eqref{param-shift-rule} via the parameter shift rules
by estimating the coefficients as, e.g., $E^{(D)}_{kl}/4$. We apply the inverse of the Hessian
to the gradient vector to update our parameters. We have proposed an optimal measurement distribution
scheme in ref.~\cite{van2021measurement} that is applicable to Hessian-based optimsiations:
the measurement costs grow with the fourth power of a regularisation parameter that we set $\eta = 0.1$.
We therefore determine the coefficients using a fixed number of shots $N_E =  10^{5}$ to populate
the Hessian matrix and we determine coefficients using $N_E = 10^{7}$ shots to populate the
gradient vector -- the latter sampling budget is comparable to the case of gradient descent,
albeit slightly higher,
such that the evolution remains stable until reaching the convergence criterion~\cite{van2021measurement}.
We use a simple Tikhonov regularisation as dicsussed in ref.~\cite{van2021measurement}. 
Note that a significant advantage of analytic descent is that it does not require a matrix inversion.

%-------------------------
\begin{figure}[tb]
	\begin{centering}
		\includegraphics[width=0.48\textwidth]{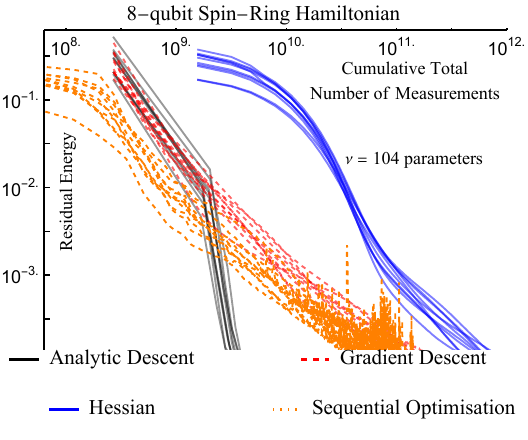}
		\caption{
			Same as in Fig.~\ref{fig2}(b) but we set the sampling rate hyper-parameter $N_E$ in the
			case of sequential optimisation the same as in case of simple gradient descent.
			In particular, we determine each coefficient, such as $E^{(B)}_k$,
			using $N_E = 10^{6.41}$ shots instead of $N_E = 10^{8}$ in the case of sequential
			optimisation. The approach becomes unstable before approaching our
			convergence criterion $\Delta E=10^{-4}$, even though initially it outperforms others.
			\label{coordesclow}
		}
	\end{centering}
\end{figure}
%-------------------------

\textbf{Sequential optimisation:} We consider the sequential optimisation techniques introduced in
refs.~\cite{PhysRevResearch.2.043158,parrish2019jacobi,ostaszewski2019quantum}. As such, we determine
and jump to the global minimum of the energy along a single parameter slice $\theta_k$ via the
update rule
\begin{equation*}
	\theta_k \rightarrow 
	\arctan \left(E^{(C)}_k - E^{(A)} , -E^{(B)}_k\right),
\end{equation*} 
as determined by coefficients  in our analytical approximation
and $\arctan(\cdot, \cdot)$ is the $2$-argument $\arctan$ function.
We set the number of shots $N_E=10^{8}$ for evaluating the coefficients, such as $E^{(C)}_k$, such that the evolution
can reach our convergence criterion. This inevitably oversamples in the early evolution and, of course,
one could adaptively set the precision. 
We stress again, however, that all other approaches would benefit from optimally setting
sampling rates throughout the evolution as already discussed in the specific case of gradient descent.
Nevertheless, we use a constant sampling rate in order to be able to compare vastly different optimisation
techniques and their convergence rates. As such, a left-to-right shift in Fig.~2 in the main text
should be viewed as an artefact of our choice.

We also note that it was necessary to choose
	a sampling rate $N_E$ larger than in case of gradient descent
	where $N_E = 10^{6.41}$ did suffice.
	We have repeated our simulation of sequential optimisation with $N_E = 10^{6.41}$
	and indeed Fig.~\ref{coordesclow} confirms that the evolution (orange dots) can become
	unstable before we approach our convergence criterion. It is also evident that sequential
	optimisation may be favourable in the early evolutions, however, note that the overall cost
	of optimisation is dominated by the later stages of the evolution.

\subsection{Analytic Descent with Quantum Natural Gradient \label{natgrad}}

Let us now show that our approximation
of ansatz circuits can be used
to obtain a classical model for computing how the quantum Fisher information matrix $\qf$ of the 
variational states $\rho(\underline{\theta}) := \Phi(\underline{\theta})\rho_0$
depends on the continuous parameters $\underline{\theta}$.
$\qf$ reduces to other notions in special cases such as the Fubini-Study metric tensor and has been used
extensively, e.g., in variational simulation or natural gradient optimisation
\cite{Li2017,samimagtime,koczor2019quantum,xiaotheory,quantumnatgrad}.
This metric tensor was first proposed in
the context of variational quantum algorithms in \cite{Li2017}
and has been used to improve convergence speed and accuracy of optimisations  as well as to avoid local minima \cite{samimagtime,koczor2019quantum,xiaotheory,quantumnatgrad,wierichs2020avoiding}.
A general approach for optimising arbitrary quantum states was proposed in \cite{koczor2019quantum}
via the quantum  Fisher information matrix;
a general approximation for noisy quantum states can be estimated via SWAP tests as $[\qf]_{mn} =  2\tr[(\partial_m \rho(\underline{\theta})) (\partial_n\rho(\underline{\theta}))]$.
Indeed, in the limit of pure states entries of this metric tensor can be
estimated using Hadamard-tests \cite{Li2017,xiaotheory, quantumnatgrad}.
We have derived the approximation of the general matrix elements $[\qf]_{mn}$
in Appendix~\ref{metric-tensor-sec}; for present purposes we need only state the leading
terms explicitly as, e.g., 
\begin{equation*}
	[\qf]_{mn} =
	\mathcal{F}_{BB} F_{BB}(\underline{\theta})
	+ \mathcal{F}_{AB} F_{AB}(\underline{\theta}) + \dots  \mathcal{O}(\sin^2\delta).
\end{equation*}
Here the multi-variate trigonometric functions, e.g., $F_{BB}(\underline{\theta})$,
can be straightforwardly computed using the previously outlined techniques.
These functions multiply the real coefficients, such  as $\mathcal{F}_{BB}$, which can
be computed from quantum-state overlaps as
\begin{align*}
	\mathcal{F}_{BB} =& 
	\tr[\rho(\tfrac{1}{2}\pi \underline{v}_k) \rho(\tfrac{1}{2}\pi \underline{v}_k)]
	+\tr[\rho(-\tfrac{1}{2}\pi \underline{v}_k) \rho(-\tfrac{1}{2}\pi \underline{v}_k)]\\
	-&\tr[\rho(-\tfrac{1}{2}\pi \underline{v}_k) \rho(\tfrac{1}{2}\pi \underline{v}_k)]
	-\tr[\rho(\tfrac{1}{2}\pi \underline{v}_k) \rho(-\tfrac{1}{2}\pi \underline{v}_k)],
\end{align*}
where $\underline{v}_k$ are basis vectors in parameter space.
These overlaps $\tr[\rho(\underline{\theta}') \rho(\underline{\theta}'')]$
correspond to variational states of shifted parameters $\underline{\theta}'$
and $\underline{\theta}''$, and can be estimated
using SWAP tests or when
the states are approximately pure as $\rho(\underline{\theta}) \approx | \psi(\underline{\theta}) \rangle \langle  \psi(\underline{\theta}) |$,
then the overlaps $|\langle \psi(\underline{\theta}') | \psi(\underline{\theta}'')\rangle |^2$
could be estimated via Hadamard tests. The latter would only require a single
copy of the state.

%-------------------------
\begin{figure*}[tb]
	\begin{centering}
		\includegraphics[width=0.7\textwidth]{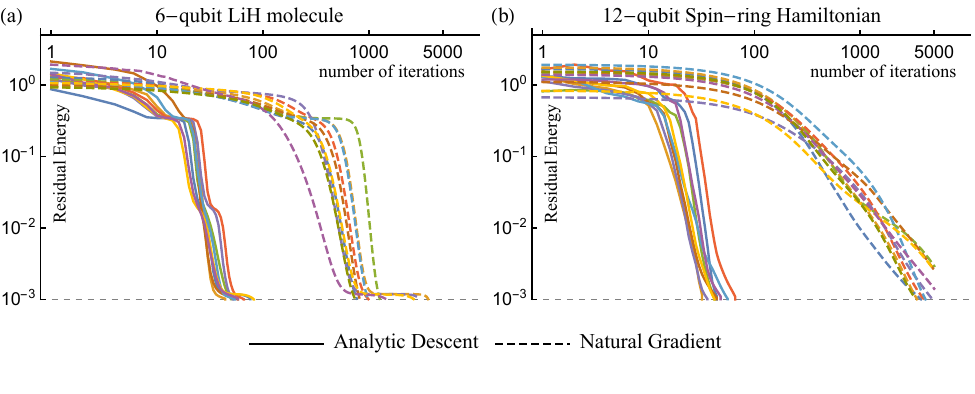}
		\caption{
			Simulated analytic descent and natural gradient in the case of Hamiltonians corresponding to (a) molecular, and (b) spin-ring systems.
			Logarithmic plots show the distance from the exact ground-state energy (Residual Energy) as a function of the iterations.
			A classical approximation of the entire energy surface is determined at each iteration step of analytic descent (solid lines) and in an internal loop we descent towards its minimum using a classical computer  (not shown here). Analytic descent (solid lines) crucially outperforms conventional natural gradient (dashed lines) and appears to increase its convergence rate (steeper slope on plots). We simulate the effect of shot noise due to finite measurements -- determining one step of analytic descent requires a factor of $2$ more measurements (included in graphs) than natural gradient. Dashed grey lines show our convergence criterion $10^{-3}$, which is comparable to chemical accuracy.
			\label{fig2_appendix}
		}
	\end{centering}
\end{figure*}
%-------------------------

Let us now apply the quantum natural gradient optimiser
whereby we multiply our classical approximation of the gradient vector with  the inverse of the metric tensor $\qf$ 
\cite{koczor2019quantum,quantumnatgrad,samimagtime}.
Although the metric tensor can be approximated classically via Eq.~\eqref{metric-tensor-eq}, we remark that its
quantum estimation cost becomes negligible in the vicinity of the optimum \cite{van2021measurement}.

We simulate the effect of shot noise in the following way.
In conventional, gradient-based optimisations one would estimate entries
of the gradient vector to a precision $\epsilon$. We set this precision
such that the relative uncertainty in the gradient vector is $10 \%$
as $(0.1 \lVert \underline{g} \rVert)^2 = \sum_{k=1}^\nu \var[g_k] $,
where $\var[g_k]$ is the variance of a single vector entry \cite{van2021measurement}. One could
distribute measurements optimally \cite{van2021measurement}, but we
set the number of measurements such that the standard deviation
of each gradient entry is $ 0.1 \lVert \underline{g} \rVert/\sqrt{\nu}$.
In order to be able to compare this to our analytic descent technique, we
determine the coefficients $E^{(B)}_k$ to the same standard deviation
$0.1 \lVert \underline{g} \rVert/\sqrt{\nu}$
and we determine all other coefficients to a proportionally inferior precision
$ 0.1 \lVert \underline{g} \rVert$.
% Recall that we can do so because the variance of
Since the variance of our classical gradient vector in Eq.~\eqref{grad_var} is dominated by the uncertainty in
$E^{(B)}_k$, 
%This way,
this way the overall number of measurements required for analytic descent is only
a factor of $2$ more than determining the gradient vector. Note that our
optimal measurement distribution strategy would of course be preferable.

Fig.~\ref{fig2_appendix}(a) shows simulation results of a LiH
Hamiltonian  of $6$ qubits. We use an ansatz circuit with
$4$ blocks and overall $78$ parameters.
We start every optimisation from a randomly selected point in parameter space
that is close to the Hartree-Fock solution.
In Fig.~\ref{fig2_appendix}(solid) we only plot the external optimisation loop
of analytic descent. 
We plot curves that correspond to analytic descent in Fig.~\ref{fig2_appendix}(solid)
such that we propagate data points by $2$ steps at every iteration -- to  reflect
their relative measurement costs.

We have used a very fine step size in the case of analytical descent, which allows us
to follow the natural gradient evolution of the parameters very smoothly ranging up to many thousands
of conventional gradient steps per a single classical optimisation procedure (one iteration in Fig.~\ref{fig2_appendix}).
This small step size has several advantages, for example, it keeps the evolution stable even when the inverse of
the ill-conditioned metric tensor $\qf$ is applied to the gradient vector.

Fig.~\ref{fig2_appendix}(b) shows simulation results of a spin-ring Hamiltonian.
We have determined the ground state of this Hamiltonian
using the previously introduced ansatz circuit, which consists of $2$ blocks and overall $84$ parameters. 
Analytic descent performs better than natural
gradient even when being far from the optimum point. The gradient in this case is
typically large and results in large steps that quickly drive away
from the reference point $\underline{\theta}_0$.
Most importantly, both Fig.~\ref{fig2_appendix}(a) and Fig.~\ref{fig2_appendix}(b) confirm our expectations and
we observe that analytic descent crucially outperforms natural gradient in the vicinity of the optimum.
In some regions -- especially when approaching the optimum --
analytic descent even appears to result in an improved convergence
rate (steeper slope in the figure).

\subsection{Details of the simulation \label{simulations}}

We use the ansatz circuit structure shown in Fig.~\ref{figansatz} in our
simulations. This consists of layers of single-qubit
$X$ and $Y$ rotations as well as layers of two-qubit
Pauli $ZZ$ gates. 

In case of Analytic Descent, at every step there is a classical optimisation
procedure involved, for which we have used the natural gradient update rule
and we aborted the internal loop when the similarity measure is low via $1-f < .5$.
We estimated the metric
tensor and inverted it using a large regularisation parameter $\eta = 0.01$ to
ensure that its measurement cost is reasonable. The step size is
$0.001$ ($0.1$) in the case of analytic descent (natural gradient).

Let us now briefly compare the measurement cost of determining $f$ to the measurement cost
of determining a single gradient vector. We first compute the variance of the estimator
\begin{equation*}
	\var[f] = \var[ \frac{ \langle \underline{\tilde{g}} | \underline{g}\rangle }
	{\lVert \underline{\tilde{g}} \rVert  \lVert \underline{g} \rVert}
	]
	\approx
	\sum_k \frac{ \tilde{g}_k^2 } {\lVert \underline{\tilde{g}} \rVert^{4} } \var[g_k]
\end{equation*}
with approximating the exact vector norm via our classical approximation's norm as 
$\lVert \underline{\tilde{g}} \rVert  \approx  \lVert \underline{g} \rVert $. For example,
assuming that $\var[g_k] = S^2$ is constant then
determining the gradient vector to a relative precision $\epsilon = r  \lVert \underline{g} \rVert$ requires
overall
$N_g = r^{-2}\,  \nu^2 S^2 / \lVert \underline{g} \rVert^2$ samples. In such a scenario we find that
the number of shots required to determine $f$ to a precision $r$ (which can be, e.g., $r=0.1$)
is given by $N_f = r^{-2} \nu S^2 / \lVert \underline{\tilde{g}} \rVert^2$ and therefore
the ratio $N_f / N_g \approx 1/ \nu$ is small in practically relevant scenarios, e.g., in our simulations the number of parameters
is large. Furthermore, we certainly do not need to query $f$ at every iteration but, e.g., at every $10$ iterations.

We consider a $6$-qubit Hamiltonian of the
LiH molecule in the following. We use an ansatz circuit with $4$ blocks and overall $78$ parameters and start the optimisation at the
vicinity of the Hartree-Fock solution. We do so by adding uniform random numbers $(-0.5, 0.5)$ to the initial parameters of the Hartree-Fock solution. The step size is
$0.001$ ($0.1$) in the case of analytic descent (natural gradient). We also determine the metric tensor at every
iteration step and regularise it with a large
$\eta=0.01$.

We also consider a 12-qubit spin-ring Hamiltonian Hamiltonian from Eq.~\eqref{hamil-appendix}:
We randomly generate $\omega_i$
and set $J=0.05$. 
We use an ansatz circuit of $2$ blocks and overall $84$ parameters. We start the optimisation from the lowest
energy computational basis state by adding uniform random numbers $(-0.5, 0.5)$ to its parameters. The step size is
$0.01$ ($0.01$) in the case of analytic descent (natural gradient).

We simulate shot noise when determining the gradient vector (in case of conventional natural gradient)
and the coefficients in Eq.~\eqref{surfaceapprox}. We do so by adding Gaussian distributed random numbers to
their exactly determined values, as discussed in the main text.

\bibliography{bibliography}

\end{document}